\documentclass[]{aastex631}
\shorttitle{AASTeX v6.3.1 Sample article}
\shortauthors{Ding et al.}
\graphicspath{{./}{figures/}}

\begin{document}

\title{Detection of Contact Binary Candidates Observed By TESS Using Autoencoder Neural Network}

\correspondingauthor{Xu Ding,ChuanJun Wang, KaiFan Ji}
\email{dingxu@ynao.ac.cn, wcj@ynao.ac.cn, jkf@ynao.ac.cn}

\author{Xu Ding}
\affiliation{Yunnan Observatories, Chinese Academy of Sciences (CAS), P.O. Box 110, 650216 Kunming, P. R. China}
\affiliation{Key Laboratory of the Structure and Evolution of Celestial Objects, Chinese Academy of Sciences, P. O. Box 110, 650216 Kunming, P. R. China}
\affiliation{Center for Astronomical Mega-Science, Chinese Academy of Sciences, 20A Datun Road, Chaoyang District, Beijing, 100012, P. R. China}

\author{ZhiMing Song}
\affiliation{School of Information, Yunnan University of Finance and Economics, Kunming, China.}
\affiliation{Yunnan Key Laboratory of Service Computing, Kunming, China.}

\author{ChuanJun Wang}
\affiliation{Yunnan Observatories, Chinese Academy of Sciences (CAS), P.O. Box 110, 650216 Kunming, P. R. China}
\affiliation{Key Laboratory of the Structure and Evolution of Celestial Objects, Chinese Academy of Sciences, P. O. Box 110, 650216 Kunming, P. R. China}
\affiliation{Center for Astronomical Mega-Science, Chinese Academy of Sciences, 20A Datun Road, Chaoyang District, Beijing, 100012, P. R. China}
\affiliation{University of the Chinese Academy of Sciences, Yuquan Road 19\#, Shijingshan Block, 100049 Beijing, P.R. China}

\author{KaiFan Ji}
\affiliation{Yunnan Observatories, Chinese Academy of Sciences (CAS), P.O. Box 110, 650216 Kunming, P. R. China}
\affiliation{Key Laboratory of the Structure and Evolution of Celestial Objects, Chinese Academy of Sciences, P. O. Box 110, 650216 Kunming, P. R. China}
\affiliation{Center for Astronomical Mega-Science, Chinese Academy of Sciences, 20A Datun Road, Chaoyang District, Beijing, 100012, P. R. China}
\affiliation{University of the Chinese Academy of Sciences, Yuquan Road 19\#, Shijingshan Block, 100049 Beijing, P.R. China}



\begin{abstract}
Contact binary may be the progenitor of a red nova that eventually produces a merger event and have a cut-off period around 0.2 days. Therefore, a large number of contact binaries is needed to search for the progenitor of red novae and to study the characteristics of short-period contact binaries. In this paper, we employ the Phoebe program to generate a large number of light curves based on the fundamental parameters of contact binaries. Using these light curves as samples, an autoencoder model is trained, which can reconstruct the light curves of contact binaries very well. When the error between the output light curve from the model and the input light curve is large, it may be due to other types of variable stars. The goodness of fit ($R^2$) between the output light curve from the model and the input light curve is calculated. Based on the thresholds for global goodness of fit ($R^2$), period, range magnitude, and local goodness of fit ($R^2$), a total of 1322 target candidates were obtained.

\end{abstract}

\keywords{Binary stars (154) --- Contact binary stars(297) --- Eclipsing binary stars(444)}


\section{Introduction} 
With the release of data from various survey telescopes, we are in an era of very large data sets in astronomy. The telescopes utilized for conducting surveys include the Zwicky Transient Facility (ZTF) \citep{Bellm+et+al+2019}, the All-Sky Automated Survey for Supernovae (ASAS-SN) \citep{Jayasinghe+et+al+2018}, the Catalina Sky Survey (CSS) \citep{Marsh+et+al+2017}, the Wide-field Infrared Survey Explorer catalog (WISE) \citep{Chen+et+al+2018}, Optical Gravitational Lensing Experiment (OGLE) \citep{Udalski+et+al+1992}, Kepler mission \citep{Borucki+et+al+2010} and Transiting Exoplanet Survey Satellite (TESS) \citep{Ricker+et+al+2015}. Eclipsing binary systems are commonly studied in astronomy, and several surveys have been conducted to identify and analyze them. Notable examples of these surveys include the Kepler \citep{prsa+et+al+2011,Kirk+et+al+2016} which identified 2878 eclipsing binaries, the OGLE \citep{Soszynski+et+al+2016} which discovered approximately 450,000 such systems, the TESS \citep{prsa+et+al+2022} which detected 4584 eclipsing binaries, the ASAS-SN \citep{Rowan+et+al+2022} which found 35,000 eclipsing binaries, and the ZTF \citep{El-Badry+et+al+2022} which identified 3879 eclipsing binaries. These large samples provide valuable insights into the properties and behaviors of eclipsing binary systems, contributing to our overall understanding of stellar evolution and the universe as a whole.

Our research in this work is specifically focused on the investigation of contact binary systems (CBs). Contact binary systems are a type of eclipsing binary star where both components fill their Roche lobes, resulting in the transfer of mass and energy between them \citep{Kopal+et+al+1959,Lucy+et+al+1968a,Lucy+et+al+1968b}. These systems are characterized by a common envelope \citep{Lucy+et+al+1979} shared by the two components and similar temperatures \citep{Kuiper+et+al+1941}. The occurrence of transients, such as Luminous Red Novae \citep{Kasliwal+et+al+2012}, has been attributed to binary systems \citep{Tylenda+et+al+2011}. To date, the observation of a confirmed merging event involving contact binaries has been limited to V1309 Sco \citep{Tylenda+et+al+2011}. Theoretical research suggests that contact binaries possess a low mass ratio cutoff and will combine into a rapidly rotating single star due to the Darwin instability \citep{Rasio+et+al+1995, Li+et+al+2006, Arbutina+et+al+2007, Arbutina+et+al+2009, Jiang+et+al+2010,
Wadhwa+et+al+2021}. The search and investigation of contact binaries with very low mass ratios is important for our understanding of the merger process and the low mass ratio limit \citep{Li+et+al+2022}. \citet{Zhang+et+al+2020} determined a new lower limit of period ($\sim 0.15$ days) by studying the correlation between the orbital periods, mass ratios, masses and radii of 365 studied targets. The possible explanations of the cut-off limit of orbital period are still under debate and observational constrains are necessary \citep{Loukaidou+et+al+2022}.

TESS provides high time resolution and high precision light curves of contact binary stars to facilitate the study of scientific research described in the previous paragraph. In order to obtain the light curves of contact binary stars
, we use an autoencoder to detect contact binaries observed by TESS in sectors 1-66. We used Phoebe \citep{prisa+et+al+2016} program to generate light curves based on the fundamental parameters of contact binary stars as samples. The reason why autoencoder can be used to detect the light curves of contact binaries is that they learn the representation of light curves of contact binaries during training and measure the difference between the input data and its own reconstruction by the reconstruction error. When the trained autoencoder model is applied to other types of variable stars, its reconstruction error will increase for that data differs significantly from the training data, indicating that it may be another type of variable star. The light curves obtained from the TESS survey were filtered using thresholds based on the global goodness of fit ($R^2$), period, range magnitude, and local goodness of fit ($R^2$). As a result, a total of 1322 targets were identified. \citet{prsa+et+al+2022} has identified 4584 eclipsing binaries from sectors 1-26 of the TESS survey. Our catalog, when compared to the catalog obtained by \citet{prsa+et+al+2022}, includes 710 contact binary candidates that are not found in their catalog.

This paper is organized as follows. In section 2, we provide a detailed description about TESS data employed in this work. In section 3, we build an autoencoder model and evaluate its reconstruction effect. In section 4, we get the results of detecting contact binary stars. In the last section, we provide our discussion and conclusions. 

\section{TESS data}
We utilized light curves obtained from the Transiting Exoplanet Survey Satellite (TESS). 
The primary object of the TESS mission is to detect and classify exoplanets through observation and analysis. TESS observes each sector, which lasts for 27.4 days, over a designated section of the sky. These long and high-precision light curves facilitate the study of periodic variable stars in them. All selected targets with a 120-second exposure time were downloaded from the Mikulski Archive for Space Telescopes (MAST) database. Light curves in sectors 1-66 observed by TESS are used to detect contact binaries. We selected light curves from pre-search Data Conditioning Simple Aperture Photometry (PDCSAP) for analysis \citep{Ricker+et+al+2015}. To determine the orbital period, the Lomb-Scargle \citep{Lomb+et+al+1976, Scargle+et+al+1982} Periodograms implemented in the Astropy \citep{Astropy+et+al+2022} package was used. 
By utilizing the Lomb-Scargle power spectrum to calculate the highest peak period, the light curve with time-flux can be transformed into a phase-magnitude light curve. The orbital period of a contact binary was set as twice the value of the highest peak period obtained through the Lomb-Scargle power spectrum analysis. To estimate the period’s uncertainty, we employ the bootstrap method \citep{Efron+et+al+1979}, which involves repeating the process with 100 measurements. To convert the flux values obtained from TESS and Phoebe into magnitudes, we utilize the following formulas.
\begin{equation}
    mag_i = -2.5 \times log10(flux_i)  
\end{equation}
\begin{equation} 
    \Delta{mag_i} = mag_i - \frac{\sum_{i=1}^n mag_i}{n}
\end{equation}

where $flux_i$ is the flux obtained from TESS and Phoebe. $\Delta{mag_i}$ is the normalized value and the input data of autoencoder network.

\vspace{2em}
\section{Method} 
We employed the Phoebe program to generate the synthetic light curves based on the fundamental parameters of contact binary. The synthetic light curves generated by Phoebe program were utilized to trian an autoencoder network model using Keras \citep{Chollet+et+al+2015}, with the aim of training a machine learning model that can reconstruct the light curves of contact binary stars accurately. The light curves generated by the trained model for contact binaries exhibit a high degree of similarity to the input data, indicating that the model reconstructs the light curves with minimal error. On the contrary, if the error between the original input and the reconstructed output using the autoencoder exceeds a certain threshold, it means that the input is anomalous data, which may be the light curves of other types of variable star.

\subsection{Autoencoder}
An autoencoder is a type of feed-forward neural network with a bottleneck shape that is primarily used for unsupervised learning tasks \citep{Ballard+et+al+1987, Baldi+et+al+2012, Hinton+et+al+2006}. It operates by compressing input data into a lower-dimensional representation, and then reconstructing the data in its original form, typically with the aim of minimizing the difference between the input and output. As a neural network architecture, an autoencoder consists of two critical components: the encoder and the decoder. The encoder reduces the dimensionality of the input data, producing a compressed representation called the code, while the decoder reconstructs the original input data from the code. This design allows the network to learn a compressed representation of high-dimensional data that can be used for various applications such as data compression, feature extraction, and object  detection from the reconstruction effect\citep{Vincent+et+al+2008}. In this work, we utilize the application of autoencoder in object detection to search for contact binary candidates.

\subsection{Samples}
In contact binary systems, the light curve is primarily influenced by several crucial parameters, including the mass ratio ($q$), the orbital inclination ($incl$), the effective temperature of the primary star ($T_1$), the effective temperature of the secondary star ($T_2$), fill-out factor ($f$), third light ratio ($l_3$), gravity-darkening coefficient ($g$), and bolometric albedos ($A$), passband, and cool spot parameters. The parameters characterizing a cool spot are commonly identified by its colatitude ($colat$) and longitude ($long$), as well as its angular radius ($radius$) and the ratio of its temperature to the intrinsic local value ($relteff$). Phoebe program is utilized to generate precise light curves of contact binaries using these fundamental parameters. The passband is set as TESS:T, which covers a broadband wavelength range of 600-1000 nm \citep{Ricker+et+al+2015}. To increase the density of our samples, we employ a random fraction approach to generate these light curves.
The mass ratio ($q$) follows a distribution within the range [0,1]. The orbital inclination ($incl$) is represented by $cos(incl)$, which is distributed within [0,1]. The fill-out factor ($f$) exhibits a distribution ranging within [0,1]. The primary star’s temperature ($T_1$) is distributed within the range of [3000K, 10000K], containing the M-A spectral types. The temperature ratio ($T_2/T_1$) was set to a Gaussian distribution with $\mu=1,\sigma=0.2$. The colatitude ($colat$) is represented by $cos(colat)$, which is distributed within [-1,1]. The longitude ($long$) is distributed in the range [0,360], the angular radius ($radius$) is distributed in the range [0,60] and the temperature to the intrinsic local value ($relteff$) is distributed in the range [0.6,1]. When considering effect temperatures below 8000K, a $g$ value of 0.32 and an $A$ value of 0.6 are set. For effect temperatures exceeding 8000K, the $g$ value is set to 1 and the $A$ value is set to 1. The atmosphere is configured as a ck2004. We also introduce a Phase Shift parameter for two reasons. Firstly, it allows us to shift the phase around 0.5, satisfying the condition for samples with a mass ratio ($q$) greater than 1. Secondly, it accounts for a certain deviation in the initial time ($T_0$), resulting in a phase offset around 0. The $\Delta{mag}$ values of Phoebe on the Y-axis are adjusted with a slight shift to better align with the actual observed TESS data. The shift value on the Y-axis was set within the range of [-0.03, 0.03]. The parameter settings involved are shown in Table 1. In Figure 1, we utilize seaborn.kdeplot \citep{Waskom+et+al+2021} to visualize the distribution of sample parameters. The number of synthesized light curves is 1,554,015.

\begin{table}
\newcommand{\tabincell}[2]{\begin{tabular}{@{}#1@{}}#2\end{tabular}} 
    \centering
	\caption{The range of the parameters of generating light curves.}\label{Table 1}
	\begin{tabular}{ccc} 
\hline \hline                        
Num     &parameter          &range                         \\
\hline
1    &$q$	                    &[0, 1]                \\
2    &$cos(incl)$	                &[0, 1]                        \\
3    &$T_1$	                &[3000K, 10000K] 	                    \\
4    &$T_2/T_1$             &($\mu=1$,$\sigma=0.2$)                  \\
5    &$f$                     &[0, 1]                           \\
6    &$l_3$                 &[0, 1]                            \\
7    &passband             & TESS:T                               \\                              
8    &atmosphere           & ck2004                        \\
9   &$g_1,g_2$             & \tabincell{c}{$g_1,g_2=0.32(T<8000K)$ \\ $g_1,g_2=1(T>=8000K)$}  \\   
10 &$A_1,A_2$  &    \tabincell{c}{$A_1,A_2=0.6(T<8000K)$ \\ $A_1,A_2=1(T>=8000K)$}\\
11 & $cos(colat)$ &[-1, 1]\\
12 & $long$ &[0, 360] \\
13 & $radius$ &[0, 60]\\
14 & $relteff$ &[0.6, 1]\\
15 & phase shift & [-0.05, 0.05]  [0.495, 0.505] \\
16 & Y-axi shift   & [-0.03, 0.03]                 \\
\hline
\end{tabular}	
\end{table}

\begin{figure}[!ht]
\begin{center}

\begin{minipage}{18cm}
	\includegraphics[width=18cm]{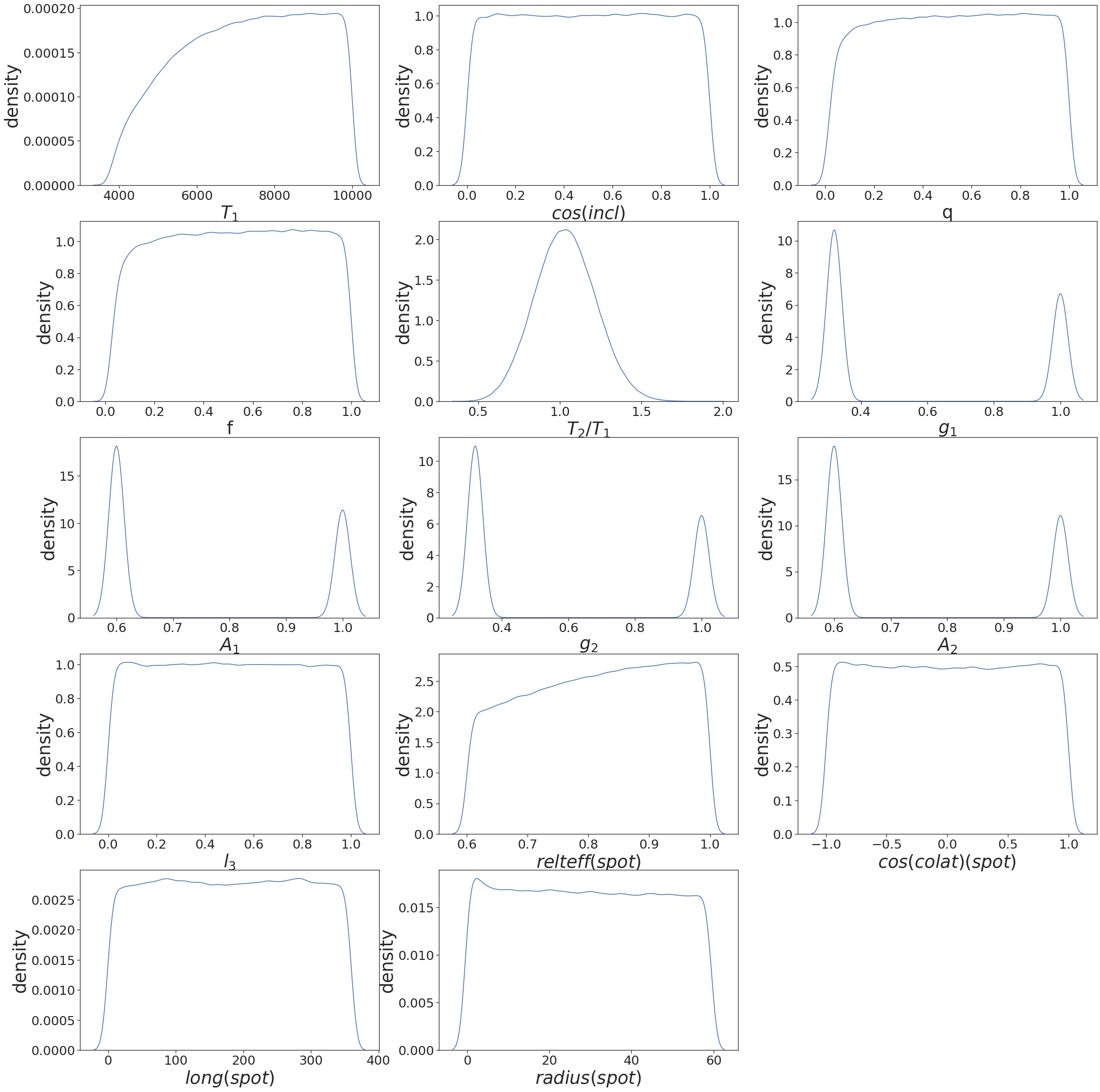}
\end{minipage}

\caption{The distribution of parameters for contact binary stars is shown.}\label{Fig1}
\end{center}
\end{figure}

\vspace{4em}
\subsection{Establishment of the autoencoder neural network} 
We conducted an analysis on 2230 light curves of non-variable stars in order to assess the noise present in TESS light curves. In Figure 2, the top panel represents the light curve of a target, known as TIC 388106759. The bottom panel in Figure 2 illustrates the distribution of the noise from 2230 targets, which is primarily centered around 0.00114.
The autoencoder neural network is predominantly composed of two key components, namely the encoder and decoder in Figure 5. The numbers in Figure 5 indicate the number of neurons per layer. $N_1$ and $N_2$ are selected as different numbers of neurons, and they are tested using the test set.
The input data consists of a light curve with 100 data points generated by Phoebe, to which Gaussian noise ($\mu,\sigma$) has been added, $\mu$ is set to 0, and the value of $\sigma$ is sampled from the Poisson distribution in Figure 2. Adding noise and applying a slight Y-axis offset is done to make the simulated data closer to the actual observed data. The output labels of the network is the original light curve without added noise, consisting of 100 points. Why choose 100 data points? We randomly generated 2707 light curves with 1000 points using Phoebe. The left panel in Figure 3 represents one of these light curves. It was interpolated to 4 points and then inverse-interpolated back to 1000 points, calculating the standard deviation of residuals compared to the original light curve. The standard deviation of residuals of the light curve in the figure is 0.2439.  The right panel in Figure 3 shows the error situation of this light curve when it is interpolated from 2 points to 200 points and then inverse-interpolated back to 1000 points. The error becomes smaller than 0.001 when this light curve is interpolated to 96 points and then inverse-interpolated back to 1000 points. Figure 4 depicts the situation where the error is less than 0.001 when 2707 light curves are interpolated to a certain number of points and then inverse-interpolated back to 1000 points. Interpolating 180 points can cover all the targets, but generating more points using Phoebe will be more time-consuming. Most targets only require interpolation within 100 points, which can meet the requirements. 
For only 16 light curves, the number of points required for interpolation exceeds 100 points. The greater the range magnitude of the light curve, the greater the number of points needed for interpolation. Interpolating the original light curve into 100 data points can effectively represent its light curve. Therefore, we utilized Phoebe to generate a 100-point phase-magnitude light curve. The number of code is 16, the main reason is that the light curve is generated by 16 variable parameters ($T_1, incl, q, T_2/T_1, f, l_3,colat, long, relteff, radius, A_1, A_2, g_1, g_2, phase \, shift, Y-axi \, shift$). The activation function is ReLU \citep{He+et+al+2015}, which is known to be effective in resolving non-linear relationships that cannot be adequately represented by linear expressions. The loss function plays a crucial role in measuring the difference between the predicted and actual values of the model. Using the variation of the loss function on the validation set as a metric helps in selecting the optimal model for subsequent training on the training set. For this study, the Mean Square Error (MSE) has been chosen as the loss function to achieve accurate predictions. The Adam \citep{Kingma+et+al+2014} optimizer was chosen for optimizing the neural network.

\begin{figure}[!ht]
\begin{center}

\begin{minipage}{18cm}
	\includegraphics[width=18cm]{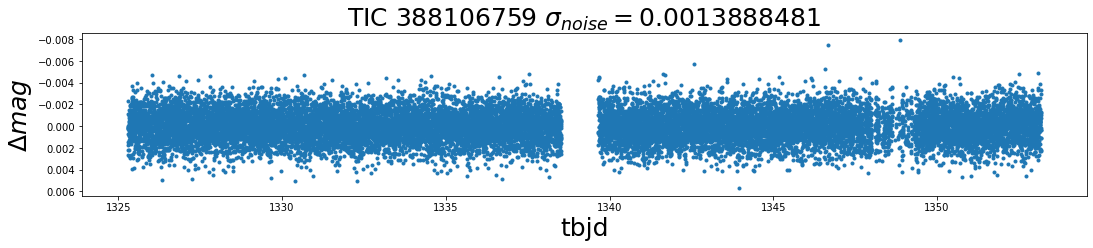}
\end{minipage}
\begin{minipage}{8cm}
	\includegraphics[width=8cm]{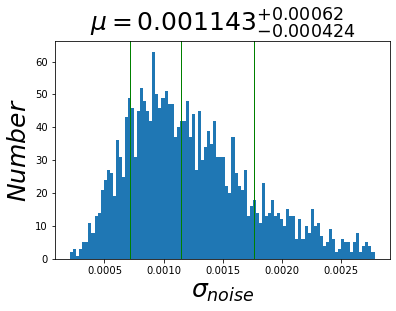}
\end{minipage}

\caption{Top: TIC 388106759 is a nonvariable star with a noise standard deviation of 0.0013888.  Bottom: noise distribution of 2230 targets similar to TIC 388106759.}\label{Fig13}
\end{center}
\end{figure}

\begin{figure}[!ht]
\begin{center}

\begin{minipage}{8cm}
	\includegraphics[width=8cm]{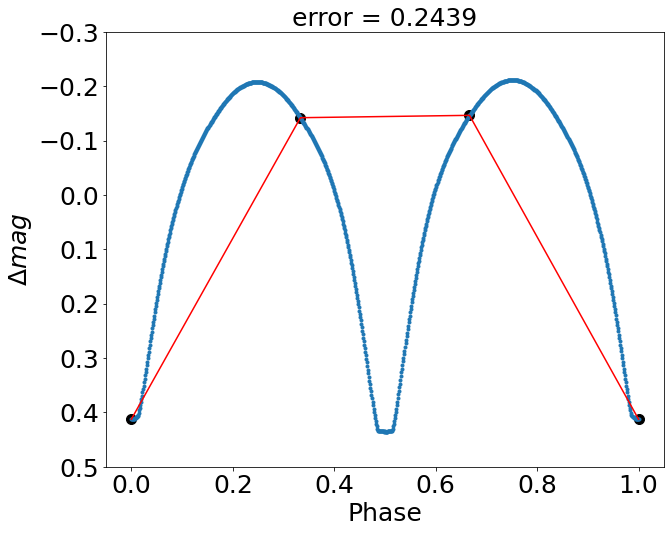}
\end{minipage}
\begin{minipage}{8cm}
	\includegraphics[width=8cm]{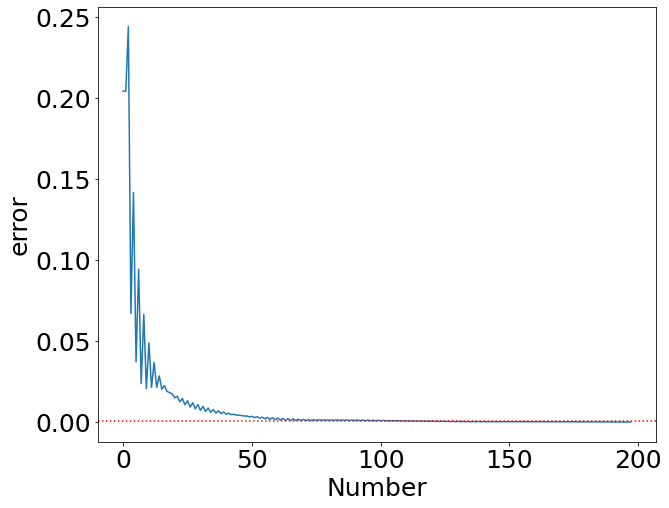}
\end{minipage}
\caption{Left: The light curve was interpolated to 4 points and then inverse-interpolated back to 1000 points, calculating the standard deviation of residuals compared to the original light curve. The standard deviation of residuals is 0.2439. Right: This figure shows the error situation of this light curve when it is interpolated from 2 points to 200 points and then inverse-interpolated back to 1000 points.}\label{Fig15}
\end{center}
\end{figure}

\begin{figure}[!ht]
\begin{center}

\begin{minipage}{8cm}
	\includegraphics[width=8cm]{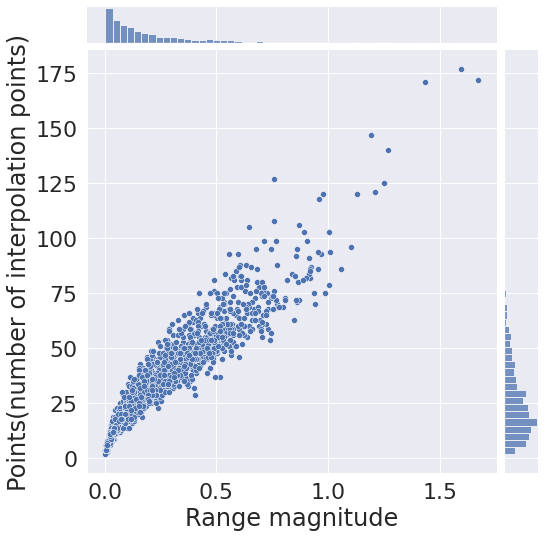}
\end{minipage}

\caption{This figure depicts the situation where the error is less than 0.001 when 2707 light curves are interpolated to a certain number of points and then inverse-interpolated back to 1000 points.}\label{Fig16}
\end{center}
\end{figure}
 
\begin{figure}[!ht]
\begin{center}

\begin{minipage}{18cm}
	\includegraphics[width=18cm]{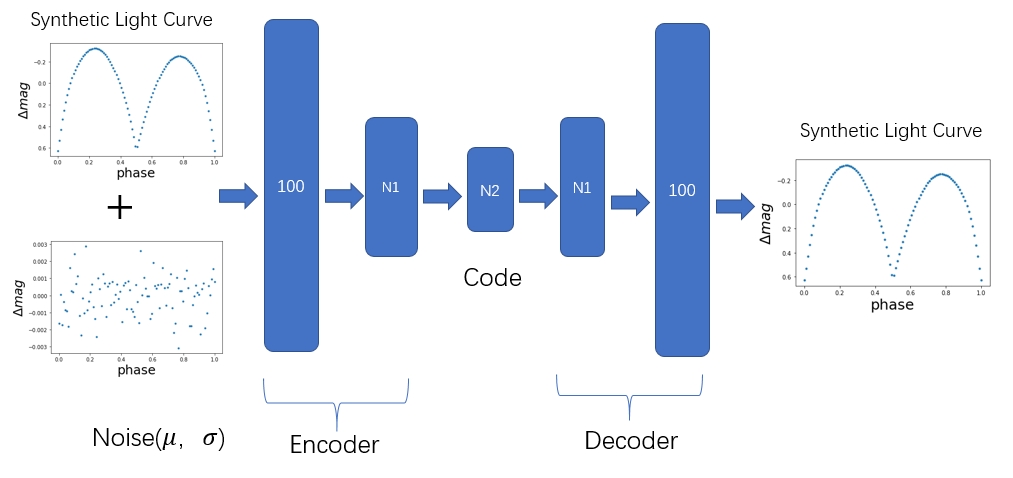}
\end{minipage}

\caption{The structure of the autoencoder neural network is shown. This structure contains the encoding structure and the decoding structure. The input of the network is a noise-added light curve of contact binary, and the output is a noise-free light curve. $\mu$ is set to 0, and the value of $\sigma$ is sampled from the Poisson distribution in Figure 2.}\label{Fig2}
\end{center}
\end{figure}

\clearpage
{
\tiny
\begin{center}
\begin{longtable}{ccccccc}
\caption{$N_1$ and $N_2$ are different numbers of neurons selected for testing, and the results are evaluated using the test set.}\label{Table 6}\\
\hline\hline                          
$N_1$     &$N_2$  &MSE  &RMSE  &MAE  &$R^2$    \\
              
\hline
\endhead
\hline
0     &  16   & $7.22\times10^{-6}$        &0.00268       &0.00131       &0.9985                    \\
70	  &  16   & $2.24\times10^{-6}$       &0.00149       &0.001023       &0.99926               \\
70    &  12   & $2.66\times10^{-6}$        &0.001632       &0.001104       &0.99913                  \\
60    &  16   & $2.27\times10^{-6}$        &0.001509       &0.0010326       &0.99925              \\
60    &  12   & $2.73\times10^{-6}$        &0.001654       &0.00111       & 0.99912                     \\ 
50    &  16   & $2.46\times10^{-6}$        &0.001571       &0.001053       &0.99921              \\
50    &  12   & $2.96\times10^{-6}$        &0.001723       &0.001144       &0.99905                      \\
40    &  16   & $2.618\times10^{-6}$        &0.001618       &0.001071       &0.99917              \\
\hline
\end{longtable}
\end{center}
}

\vspace{2em}
\subsection{Reconstruction effect of autoencoder model}
We choose 150,000 light curves as a test set, and this test set is used to evaluate the effect of the final reconstructed light curve. We use the Early Stopping technique to stop training when the model's performance on the validation set doesn't improve within 500 training epochs and obtain the model with the lowest loss value throughout the entire training process. The reconstructed light curves are generated by the autoencoder model, and the input is a light curve with Gaussian noise. Training once takes several days, so it is not feasible for us to use a very dense grid for parameter selection. Table 2 presents the testing results about MSE (Mean Squared Error), RMSE (Root Mean Squared Error), MAE (Mean Absolute Error) and goodness of fit ($R^2$), showing that adding an additional layer with $N_1$ significantly improves the network performance compared to not having $N_1$. The network performance decreases significantly for $N_2=12$ compared to $N_2=16$, therefore $N_2=16$ is chosen.
Furthermore, by considering both network simplification and performance, it is observed that varying the $N_1$ quantity does not lead to significant differences in network performance. Therefore, for practical purposes, a lesser number of $N_1=50$ was chosen, even though this setting may not be optimal. Nonetheless, it meets the requirements in practice. The left panel of Figure 6 shows the reconstructed effects of one light curve. The blue dots indicate the synthetic light curve with Gaussian noise added. The orange light curve represents the reconstructed light curve, which is generated by the autoencoder model based on the input blue light curve points. The goodness of fit ($R^2$) between the orange light curve and the input blue light curve is 0.9995. The distribution of the goodness of fit ($R^2$) between the reconstructed light curve and the input light curve with noise was calculated. The goodness of fit ($R^2$) for 150,000 light curves is mainly distributed at $0.999^{+0.00085}_{-0.02}$ in the right panel of Figure 6.
Observing Figure 6 reveals the presence of outliers exhibiting a poor goodness of fit ($R^2$), while Figure 7 sheds light on the predominant relationship between the goodness of fit ($R^2$) and the range magnitude. The smaller the range magnitude in the light curve, the greater the degree to which it is submerged in noise within the light curve. For most targets, when the range magnitude is greater than 0.1, the goodness of fit ($R^2$) is greater than 0.99. For most targets, when the range magnitude lies within the range of 0.03 and 0.1, the goodness of fit ($R^2$) surpasses the value predicted by the given formula.
\begin{equation}
    R^2 = -16x^2+3.7x+0.78
\end{equation}

where $x$ is the range magnitude.

\begin{figure}[!ht]
\begin{center}

\begin{minipage}{8cm}
	\includegraphics[width=8cm]{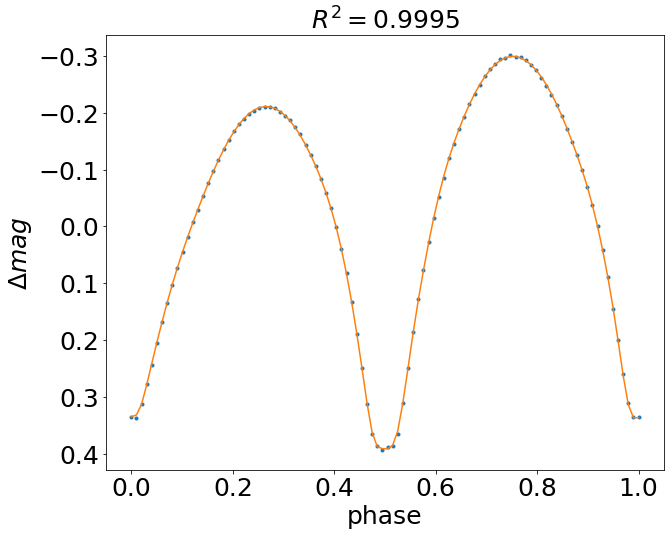}
\end{minipage}
\begin{minipage}{8cm}
	\includegraphics[width=8cm]{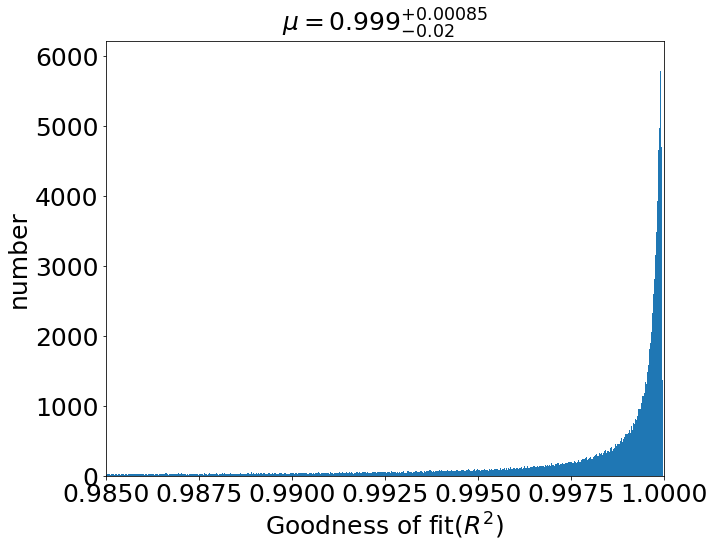}
\end{minipage}

\caption{Left: The blue dots are the light curve generated by Phoebe with Gaussian noise added. The orange light curve is generated by the trained autoencoder model using the blue dots as inputs. Right: The distribution of the goodness of fit ($R^2$) between the 150,000 input light curves and the responding output light curves is shown.}\label{Fig3}
\end{center}
\end{figure}

\begin{figure}[!ht]
\begin{center}

\begin{minipage}{12cm}
	\includegraphics[width=12cm]{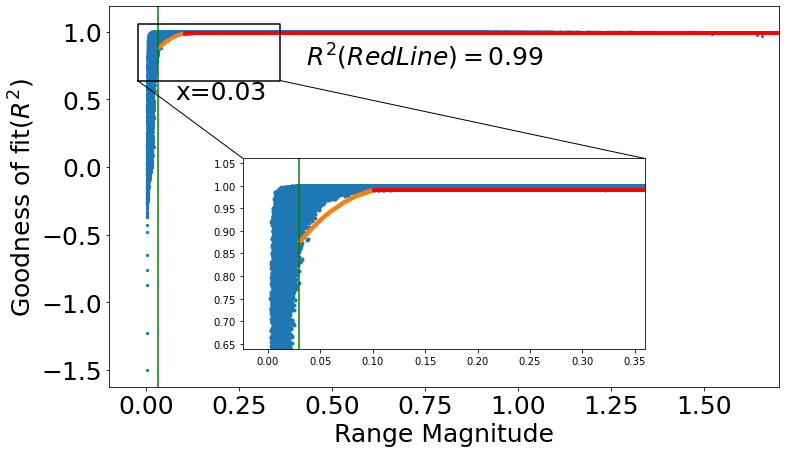}
\end{minipage}
\caption{Relationship between goodness-of-fit ($R^2$) and the range magnitude. When the range magnitude exceeds 0.1, the goodness of fit ($R^2$) value is greater than 0.99 and is represented by a red line. When the range magnitude falls between 0.03 and 0.1, the goodness of fit ($R^2$) surpasses a value calculated by formula 3, which is depicted by an orange line.}\label{Fig23}
\end{center}
\end{figure}


\clearpage
\section{Results}

\subsection{Contact binary candidates from 1-66 sectors}
 Light curves with phase-magnitude from TESS survey are fed into the autoencoder for reconstruction. We employ a three-step approach to primarily select candidates for contact binary systems. Firstly, based on figure 7, we establish a threshold value for the goodness of fit ($R^2$) which is 0.99 to identify and exclude targets exhibiting notable irregularities. We select targets whose overall light curve exhibits a range magnitude greater than 0.1. 
 The confidence level of the contact binary candidates with low range magnitude is relatively low. Variable stars with low range magnitude may encompass ellipsoidal variables, rotating variables, and pulsating variables, among others. Subsequently, we utilize a threshold value of 1.2 days for the period, in order to effectively filter out candidates with favorable fits that may have a higher probability of not being contact binary systems. The reason for excluding long-period variable stars is that they are highly likely not to be contact binary stars. Lastly, we employ a sliding window approach to calculate local goodness of fit ($R^2$), further refining the selection of targets.
 
 We can filter prominently anomalous targets by setting a threshold for goodness of fit ($R^2$). For targets that show obvious differences in their light curve compared to contact binary stars, it is easy to screen them out using the goodness of fit ($R^2$) value. In this paper, we set the threshold of $R^2$ to 0.99, which can filter out some obvious abnormal targets, as shown in Figure 8. There is a non-variable stars among them, such as TIC 25064377. The reconstruction effects of the eclipsing binaries are shown, such as TIC 25155310, TIC 33715938 and TIC 358457398. TIC 25155310 is showing planetary transits. The reconstructed effects of pulsating variable stars are shown, such as TIC 234518883, TIC 8935864, TIC 51991595 and TIC 126659093. TIC 234518883 and TIC 89358641 are the Ab-type RR Lyrae variable stars. TIC 51991595 and TIC 126659093 are the $\delta$ Scuti type stars. In Figure 8, the blue dots represent the original light curves, the orange curves represent the interpolated lines, and the green dots represent the light curves reconstructed by the autoencoder model. The light curves of these types are not well reconstructed by the autoencoder model. The goodness of fit ($R^2$) makes it possible to remove these variable stars as contact binary candidates.

 \citet{Latkovi+et+al+2021} have compiled statistics for 700 individually studied W UMa stars. The period of contact binary stars is mainly concentrated within 1.2 days \citep{Latkovi+et+al+2021}. By setting the threshold of the period to 1.2, we can filter out some targets that are well-fitted but not the contact binaries in Figure 9. 
 Please note that the 1.2 day period is considered from a statistical perspective, and the restriction of this period is not a very strict criterion. However, this limitation condition is effective in filtering out the majority of long-period abnormal targets. We have examined the variable star types presented in Figure 9 from Simbad \footnote{https://simbad.harvard.edu/simbad/} and the International Variable Star Index (VSX) \footnote{https://www.aavso.org/vsx/} website. The star identified as TIC 326630342 is classified as a semidetached binary system, specifically known as V375 Cassiopeia \citep{LiQian+et+al+2022}. TIC 13123877 corresponds to a classical Cepheid variable star. Additionally, TIC 22112755 is categorized as a rotating variable star. TIC 21540586, TIC 24137126, TIC 28828678 and TIC 29953651 are specifically identified as an Alpha2 CVn variable star. TIC 21120520 appears to be a binary star system, possibly classified as either a detached binary or a semidetached binary. TIC 13123877, TIC 21540586, TIC 22112755, TIC 24137126, TIC 28828678 and TIC 29953651 can also be filtered out by using a range magnitude threshold of 0.1.

After applying thresholds on the goodness of fit ($R^2$), range magnitude and period, a sliding window approach is employed to calculate local goodness of fit ($R^2$), further filtering the targets with poor local fitting. Windows that are too small are easily affected by local outliers, while windows that are too large smooth out too many points, failing to check the local fitting effect. The window size is chosen as 0.15 of the overall length. 
The local goodness of fit ($R^2$) is also approximately adopting the threshold of the global goodness of fit ($R^2$). The filtering rules follow the illustration in Figure 7. When the local variation exceeds 0.1 magnitudes, the local goodness of fit ($R^2$) is greater than 0.99. When it is greater than 0.03 and less than 0.1 magnitudes, the local goodness of fit ($R^2$) is greater than Formula 3. When it is less than 0.03, the local goodness of fit ($R^2$), was set greater than 0. In Figure 10, the global goodness of fit ($R^2$) of these targets is greater than 0.99, but the local goodness of fit ($R^2$) can exclude these targets. The local conditions are represented by $Range.mag_{red}$, $R^2_{red}$ and $\sigma_{red}$ in figure 10. $\sigma_{red}$ represents the standard deviation of local residuals. TIC 26261539, TIC 126321357, TIC 96051791 and TIC 103024917 are C-type RR Lyrae variable stars. TIC 123416563, TIC 24586751, TIC 34145330 and TIC 70253276 remains uncertain, as they exhibit properties consistent with being either a detached binary or a semi-detached binary system.

We list the reconstruction results for the eight light curves of contact binary candidate, as shown in Figure 11. Magnetic activity is often observed for late-type spectroscopic contact binaries and may lead to O’Connell effect \citep{Connell+et+al+1951, Milone+et+al+1968}. The O'Connell effect refers to the observation that the maxima in the light curve of eclipsing binary stars are not of equal brightness. The light curves of eight targets are well reconstructed, including several of them with O’Connell effect. For light curves with minor phase shifts, it can also fit very well, such as TIC 627436 and TIC 4164713. Light curves with scattered outliers causing slight Y-axis shifts can also be fitted very well, such as TIC 4430704 and TIC 11480757.  We provide a catalog that contains the contents of target name, right ascension (RA), declination (DEC), orbital period, period error ($\sigma_{period}$), Range magnitude, sectors observed and goodness of fit ($R^2$) in Table 3 \footnote{https://github.com/dingxu6207/TECC}. There are a total of 1322 targets in this Table. All TESS eclipsing binaries (EBs) have Gaia DR2 \citep{Gaia+et+al+2018} designations in Simbad, which can be queried directly using astroquery \citep{Astropy+et+al+2022}. A total of 1175 targets could be successfully matched and have temperature information. The period, goodness of fit($R^2$), and temperature distributions are shown in Figure 12. The distribution of periods is mainly $\mu = 0.4165^{+0.293}_{-0.106}$ day, the distribution of temperatures is $\mu = 6016^{+1071}_{-753}$K, and the distribution of goodness of fit ($R^2$) is $\mu = 0.9996^{+0.0002}_{-0.001}$. \citet{prsa+et+al+2022} obtained 4584 eclipsing binaries from 1-26 sectors from TESS survey. Compared with the catalog obtained from \citet{prsa+et+al+2022}, our catalog has 710 targets that are not present in this catalog. Figure 13 shows ten candidate targets among them.

{
\tiny
\begin{center}
\begin{longtable}{ccccccccc}
\caption{Catalog of contact binary star candidates}\label{Table 61}\\
\hline\hline                          
Name (TIC)     &RA  &DEC  &Period(day)  &$\sigma_{period}$(day)  &Range.magnitude  &Sectors  &$R^2$    \\
              
\hline
\endhead
\hline
627436      &  73.04731162   &  -25.1940849   &  0.57937     &0.000122        &0.49183       &32, 5                    &0.9997                   \\
4164713	    &  114.625919    & 33.711681      &  0.27492     &0.000059        &0.1776        &20, 45, 47, 46, 60, 44   &0.9998                \\
4430704     &  116.8675056   & 33.44015891    &  0.36776     &0.000092        &0.346539      &45, 46, 44               &0.9993                  \\
5260642     &  269.7025746   & 13.49618949    &  0.34493     &0.000084        &0.742067      &53                       &0.9998              \\
82120025    &  197.6342247   & -4.159040172   &  0.31131     &0.000073        &0.390794      &46                       &0.9999                     \\ 
5795305     &  348.750395    & -15.04677279   &  0.37369     &0.000072        &0.24713       &2, 42                    &0.9998              \\
11480757    &  303.9830976   & 37.45431286    &  0.40389     &0.000095        &0.26744       &41, 55, 15               &0.9992                      \\
17563811    &  243.8341211   & 35.7072952     &  0.36116     &0.000188        &0.292478      &24, 25                   &0.9997              \\
...         &...    &...        &...        &...        &... &... &...                 \\
\hline
\end{longtable}
\end{center}
}

\clearpage
\begin{figure}[!ht]
\begin{center}
\begin{minipage}{6.5cm}
	\includegraphics[width=6.5cm]{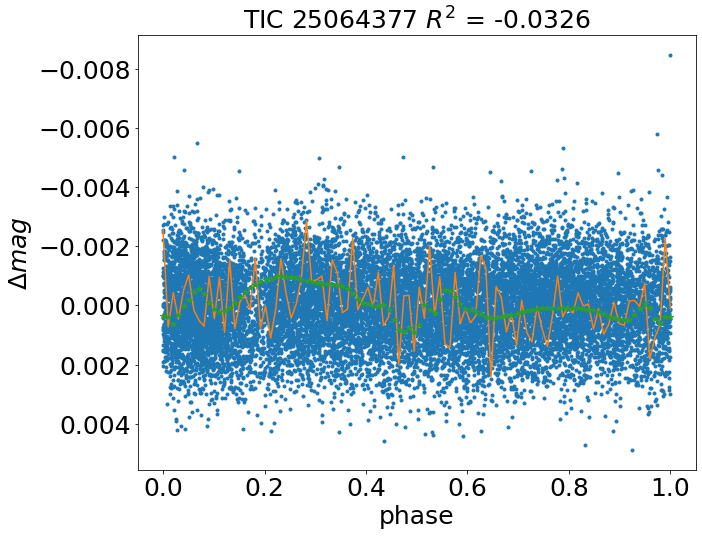}
\end{minipage}
\begin{minipage}{6.5cm}
	\includegraphics[width=6.5cm]{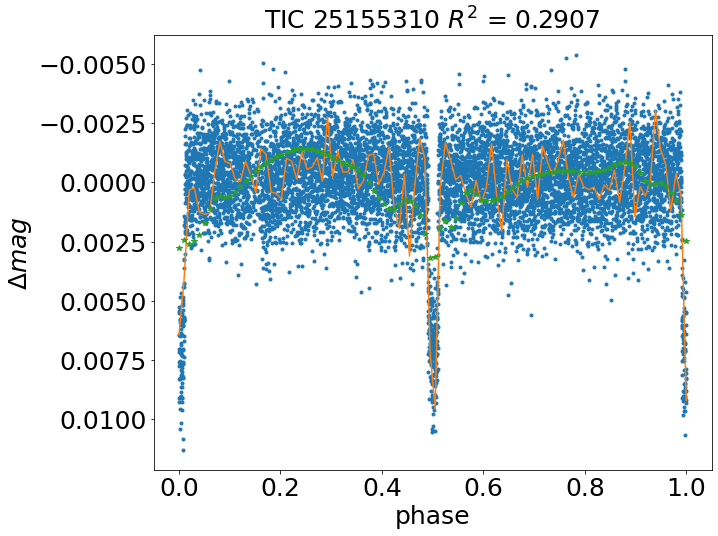}
\end{minipage}
\begin{minipage}{6.5cm}
	\includegraphics[width=6.5cm]{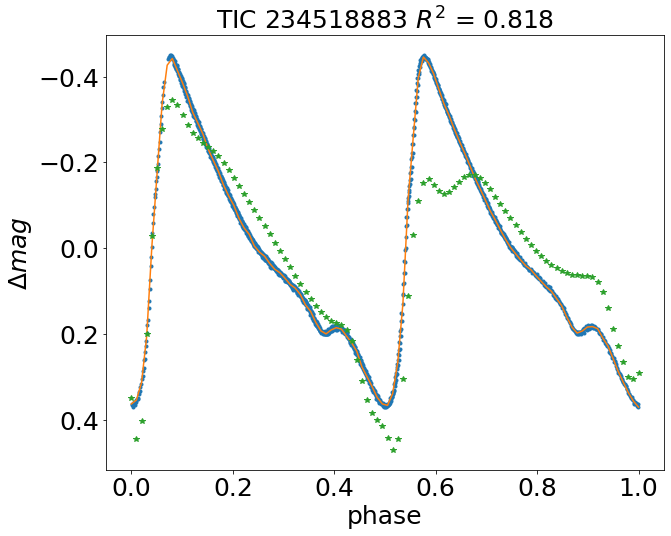}
\end{minipage}
\begin{minipage}{6.5cm}
	\includegraphics[width=6.5cm]{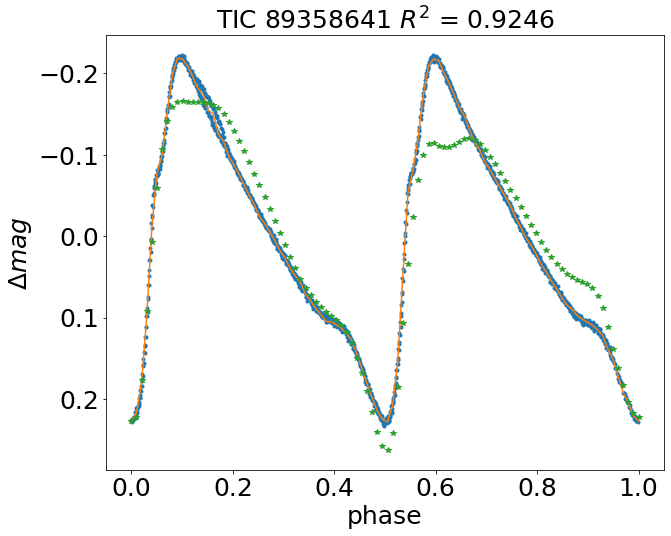}
\end{minipage}
\begin{minipage}{6.5cm}
	\includegraphics[width=6.5cm]{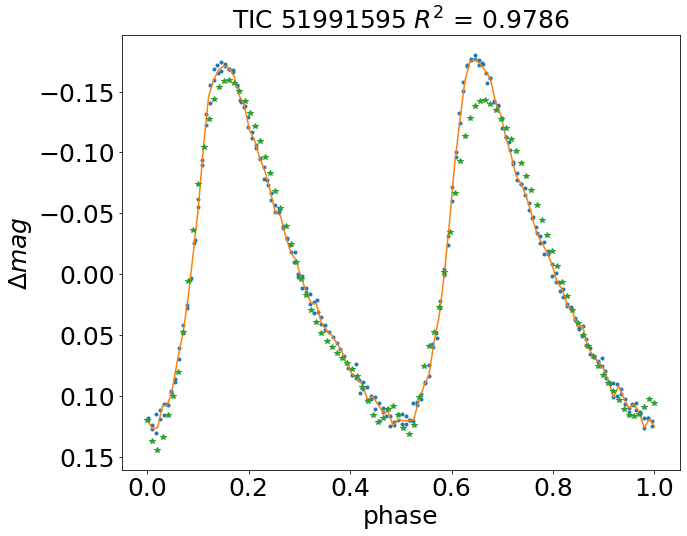}
\end{minipage}
\begin{minipage}{6.5cm}
	\includegraphics[width=6.5cm]{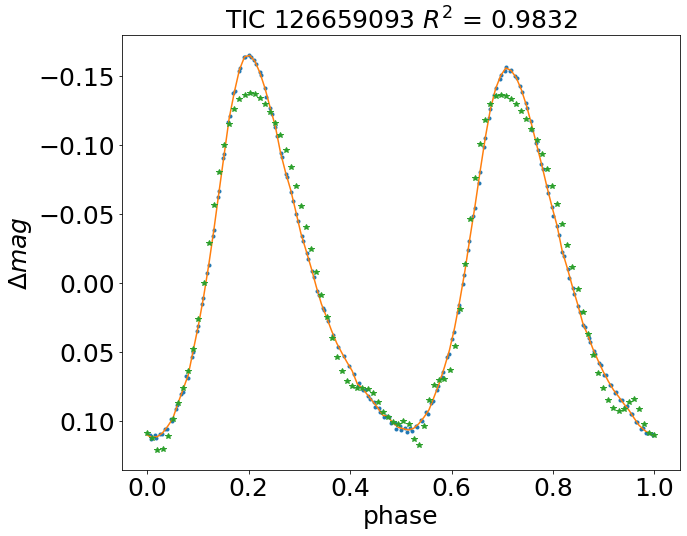}
\end{minipage}
\begin{minipage}{6.5cm}
	\includegraphics[width=6.5cm]{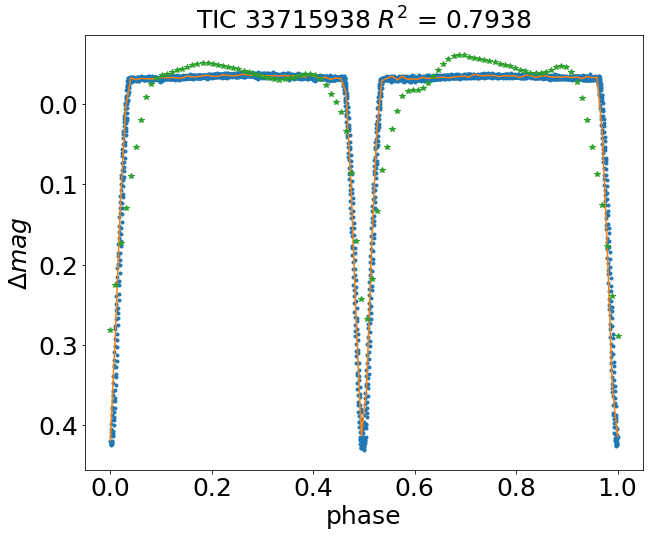}
\end{minipage}
\begin{minipage}{6.5cm}
	\includegraphics[width=6.5cm]{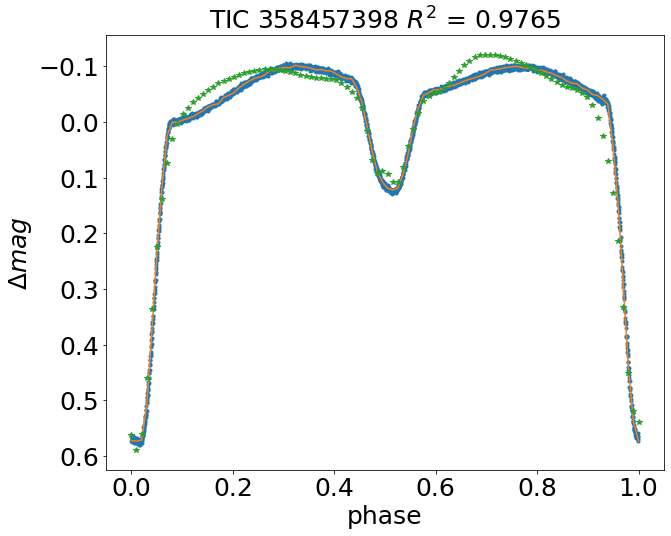}
\end{minipage}

\caption{Light curves of other types can not be reconstructed well. The blue dots indicate the original light curve, the orange lines indicate the interpolated curves, and the green dots indicate the reconstructed light curves.}\label{Fig76}
\end{center}
\end{figure}

\clearpage
\begin{figure}[!ht]
\begin{center}
\begin{minipage}{6.5cm}
	\includegraphics[width=6.5cm]{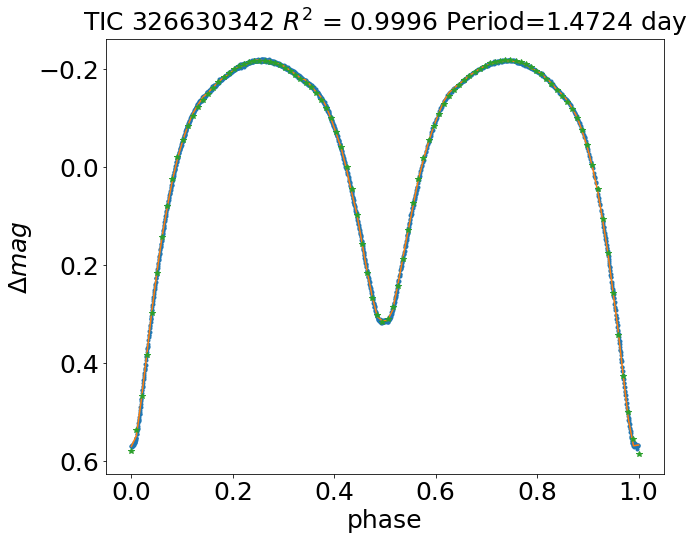}
\end{minipage}
\begin{minipage}{6.5cm}
	\includegraphics[width=6.5cm]{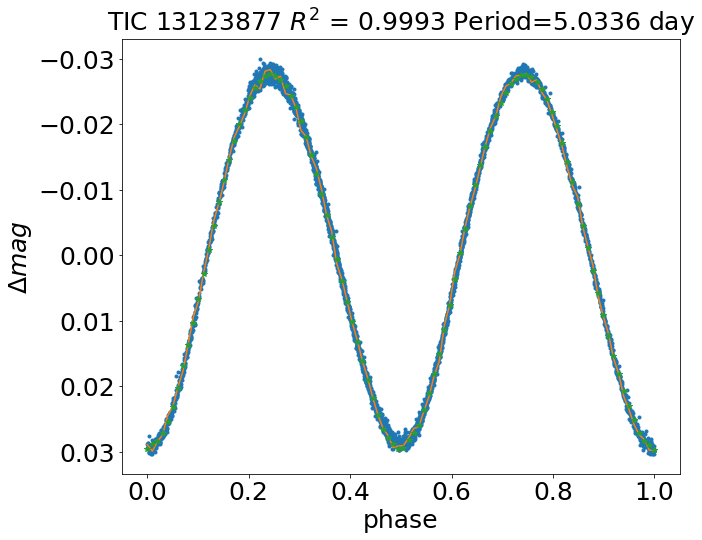}
\end{minipage}
\begin{minipage}{6.5cm}
	\includegraphics[width=6.5cm]{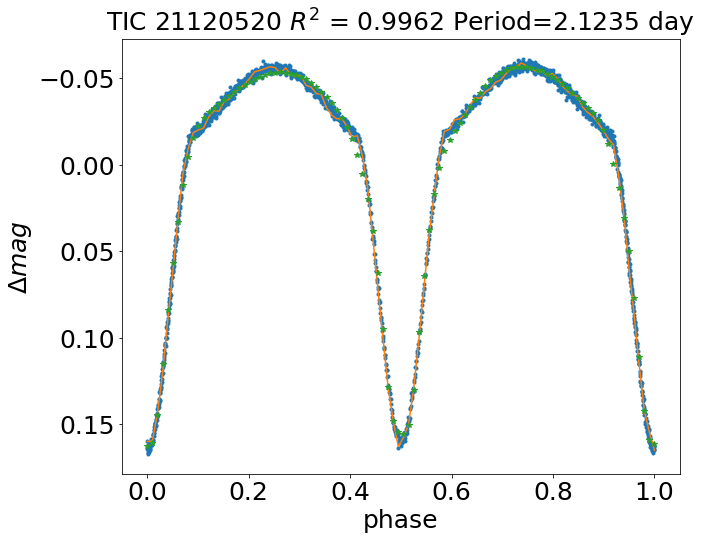}
\end{minipage}
\begin{minipage}{6.5cm}
	\includegraphics[width=6.5cm]{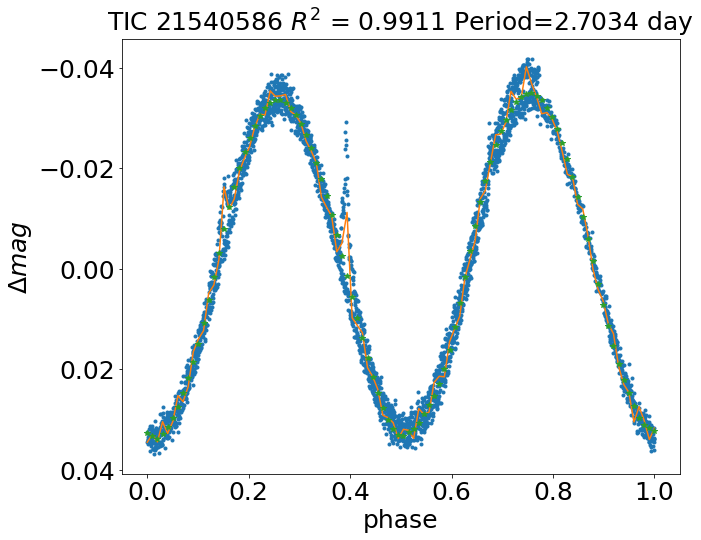}
\end{minipage}
\begin{minipage}{6.5cm}
	\includegraphics[width=6.5cm]{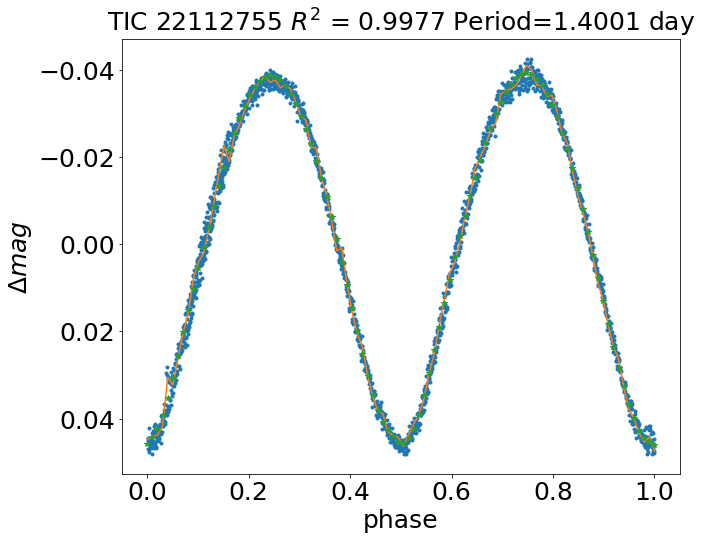}
\end{minipage}
\begin{minipage}{6.5cm}
	\includegraphics[width=6.5cm]{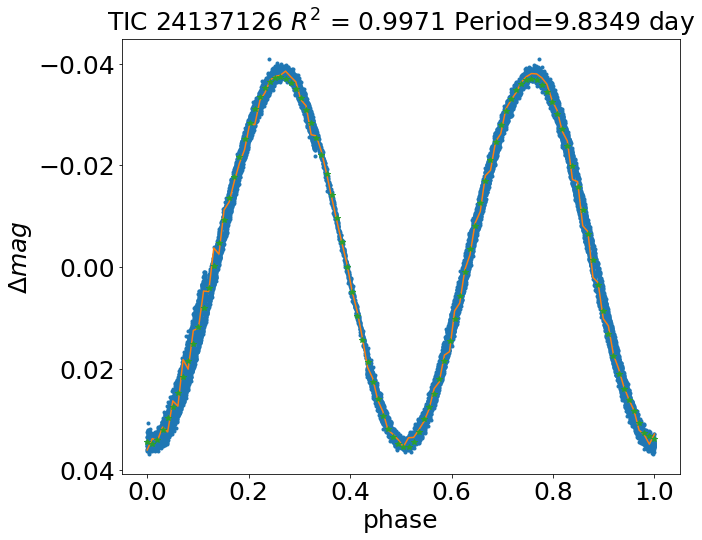}
\end{minipage}
\begin{minipage}{6.5cm}
	\includegraphics[width=6.5cm]{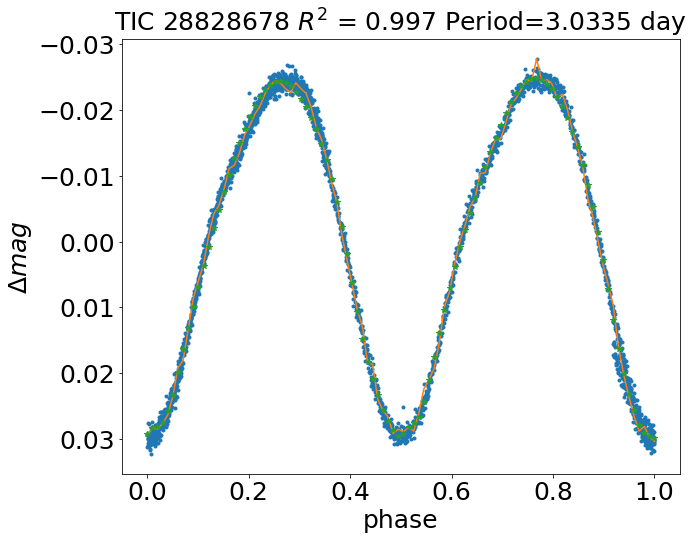}
\end{minipage}
\begin{minipage}{6.5cm}
	\includegraphics[width=6.5cm]{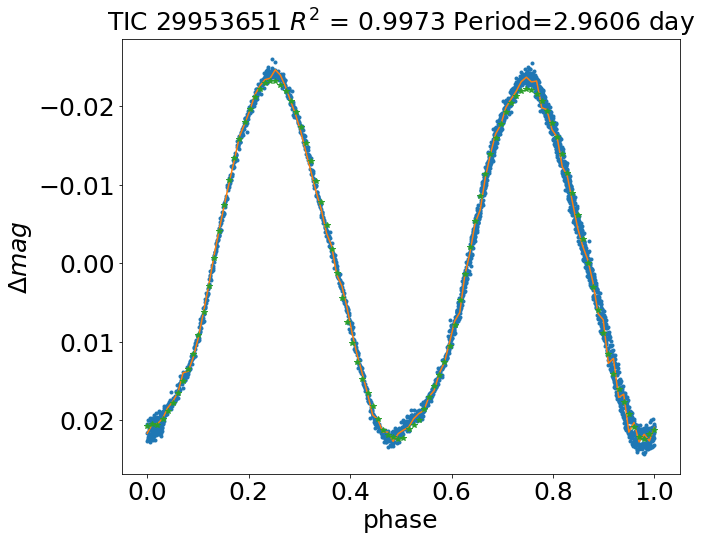}
\end{minipage}

\caption{Targets with a period greater than 1.2 days are excluded. The blue dots indicate the original light curve, the orange lines indicate the interpolated curves, and the green dots indicate the reconstructed light curves.}\label{Fig7}
\end{center}
\end{figure}

\clearpage
\begin{figure}[!ht]
\begin{center}
\begin{minipage}{7cm}
	\includegraphics[width=7cm]{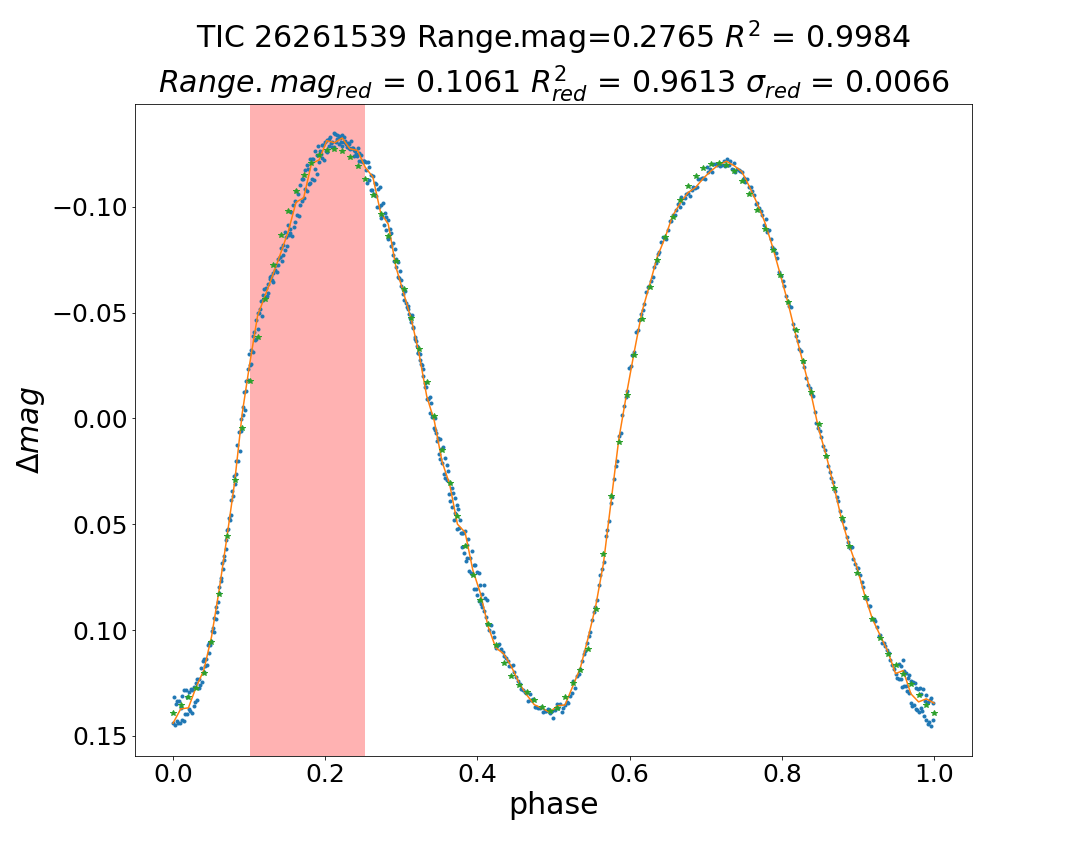}
\end{minipage}
\begin{minipage}{7cm}
	\includegraphics[width=7cm]{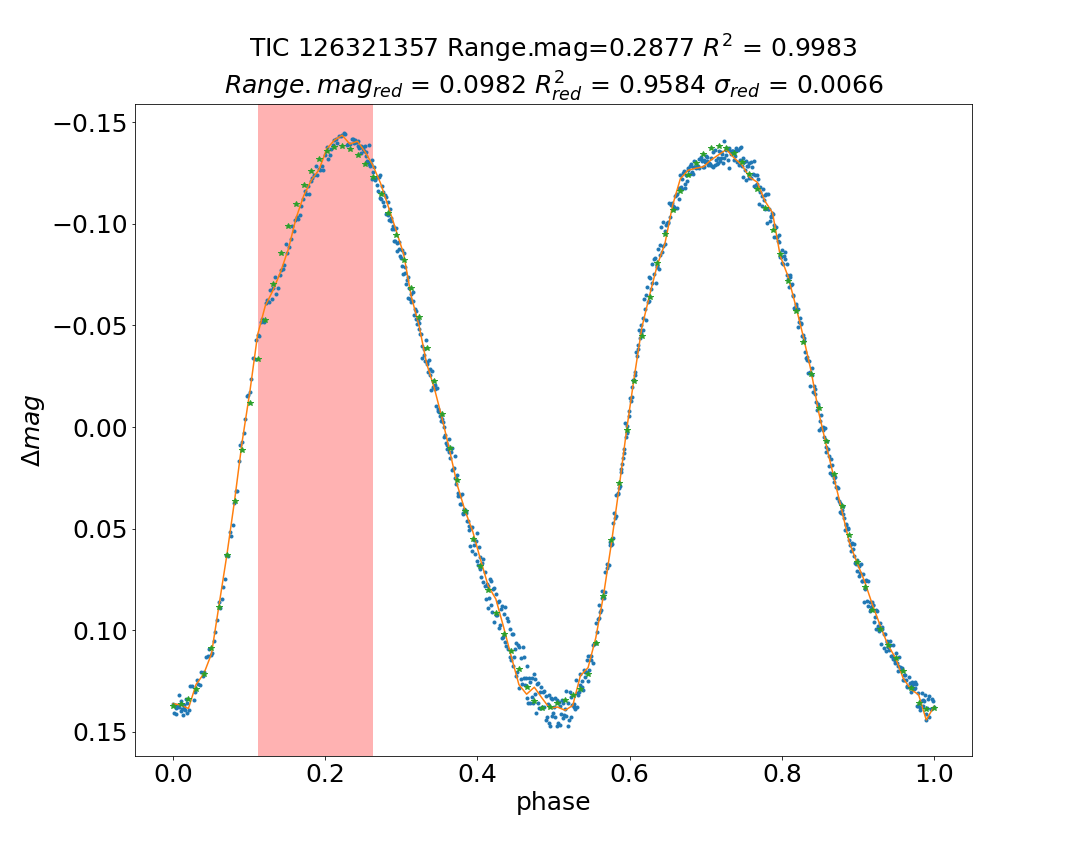}
\end{minipage}
\begin{minipage}{7cm}
	\includegraphics[width=7cm]{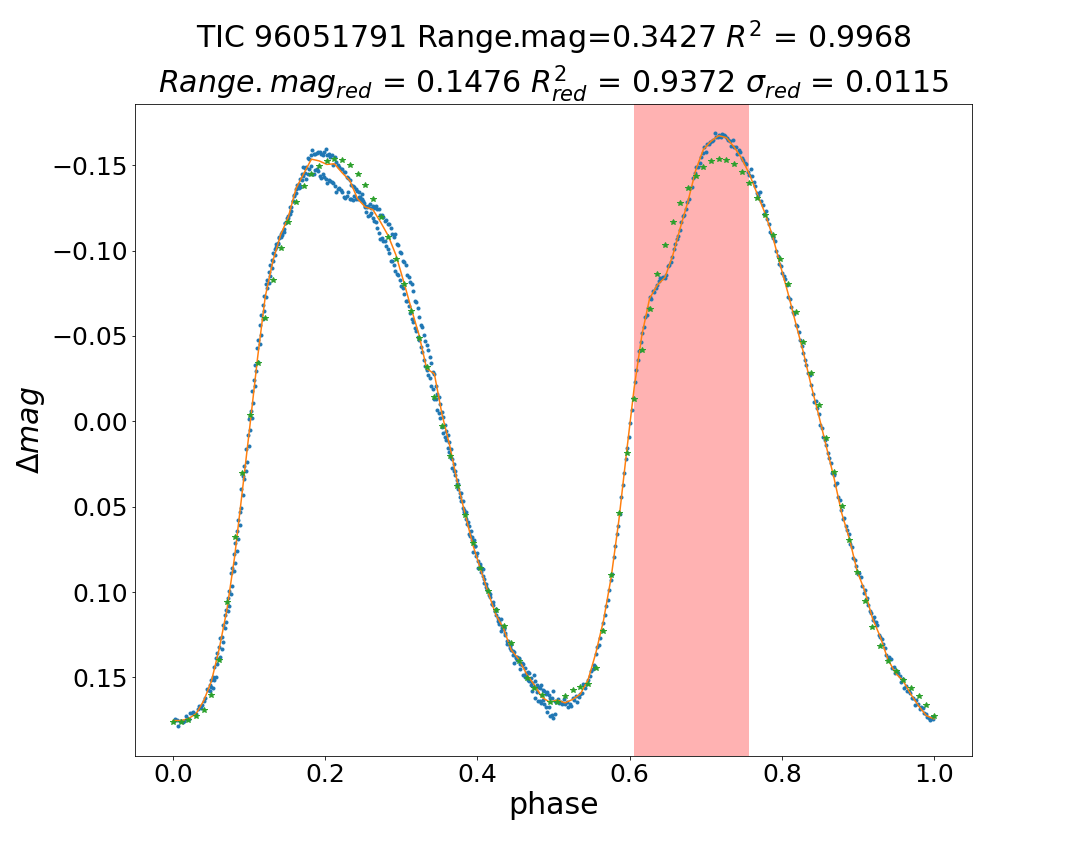}
\end{minipage}
\begin{minipage}{7cm}
	\includegraphics[width=7cm]{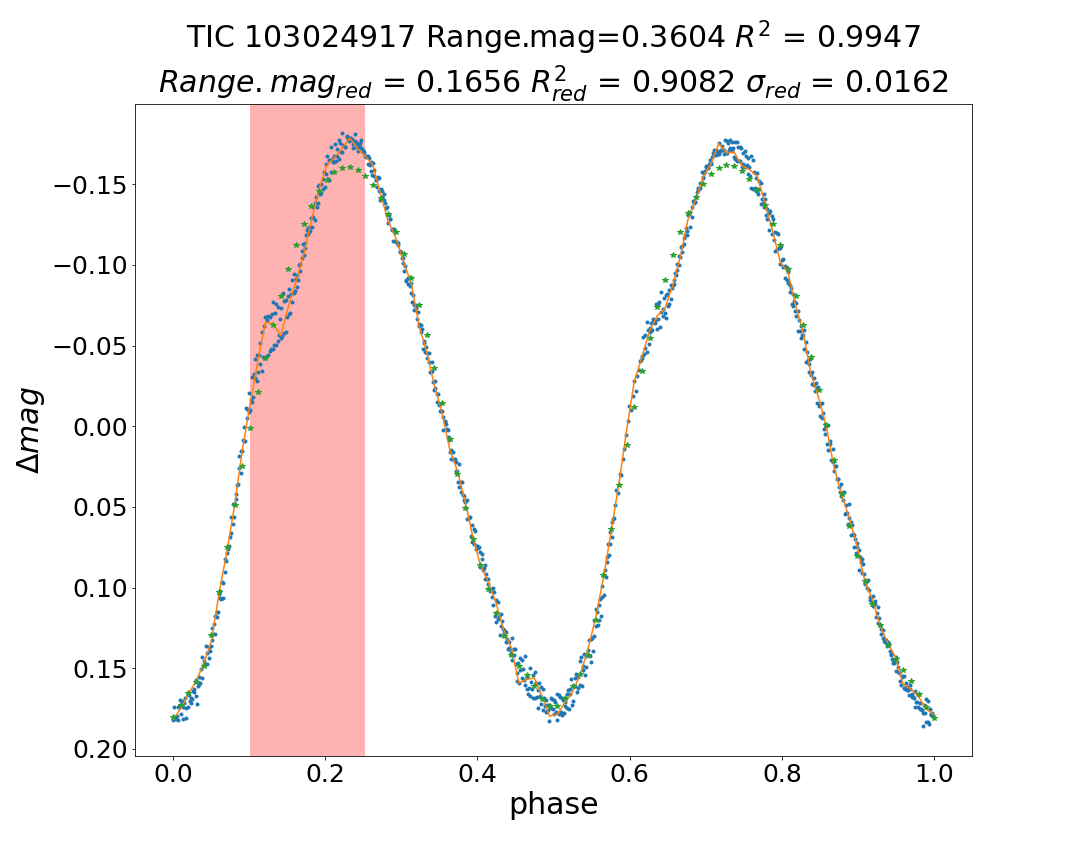}
\end{minipage}
\begin{minipage}{7cm}
	\includegraphics[width=7cm]{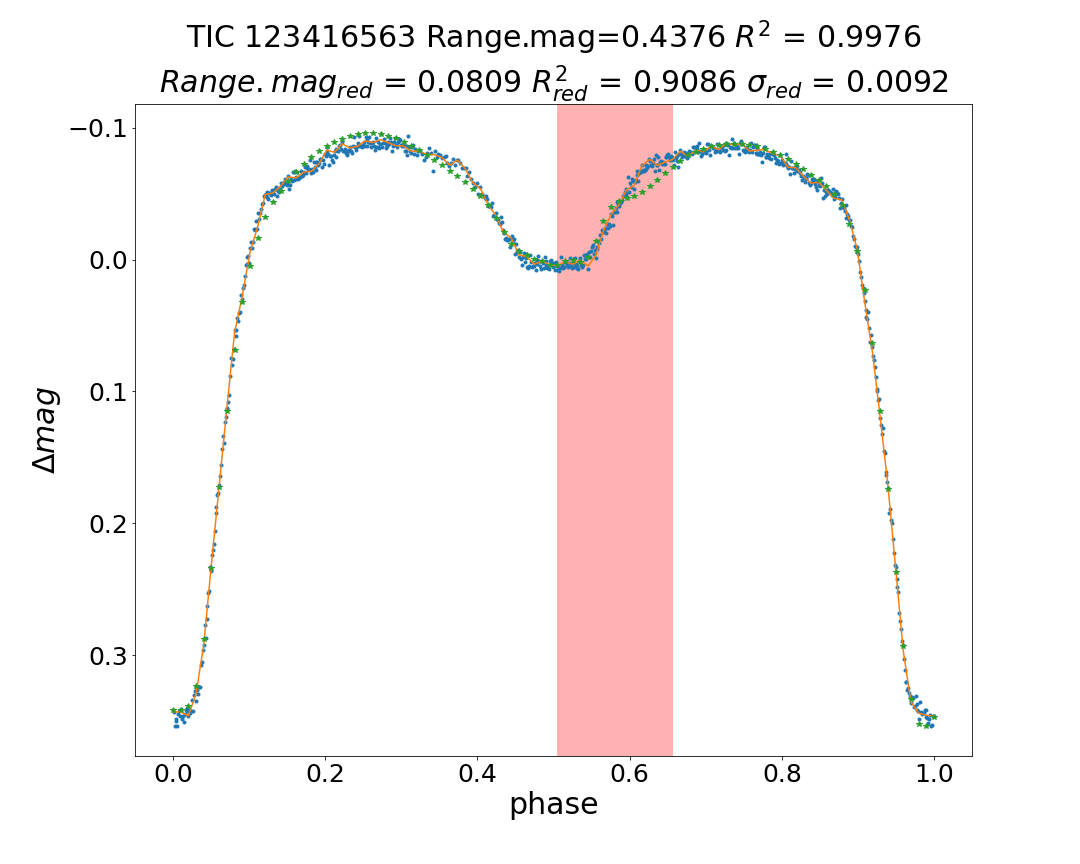}
\end{minipage}
\begin{minipage}{7cm}
	\includegraphics[width=7cm]{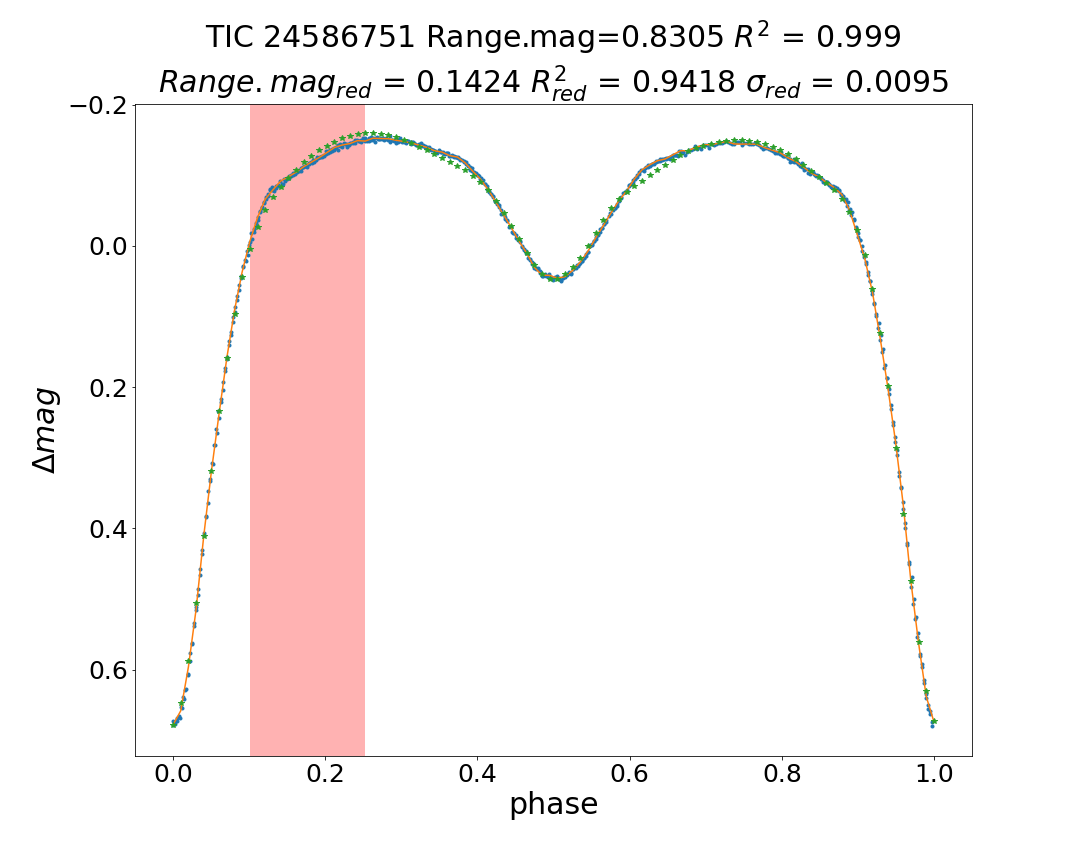}
\end{minipage}
\begin{minipage}{7cm}
	\includegraphics[width=7cm]{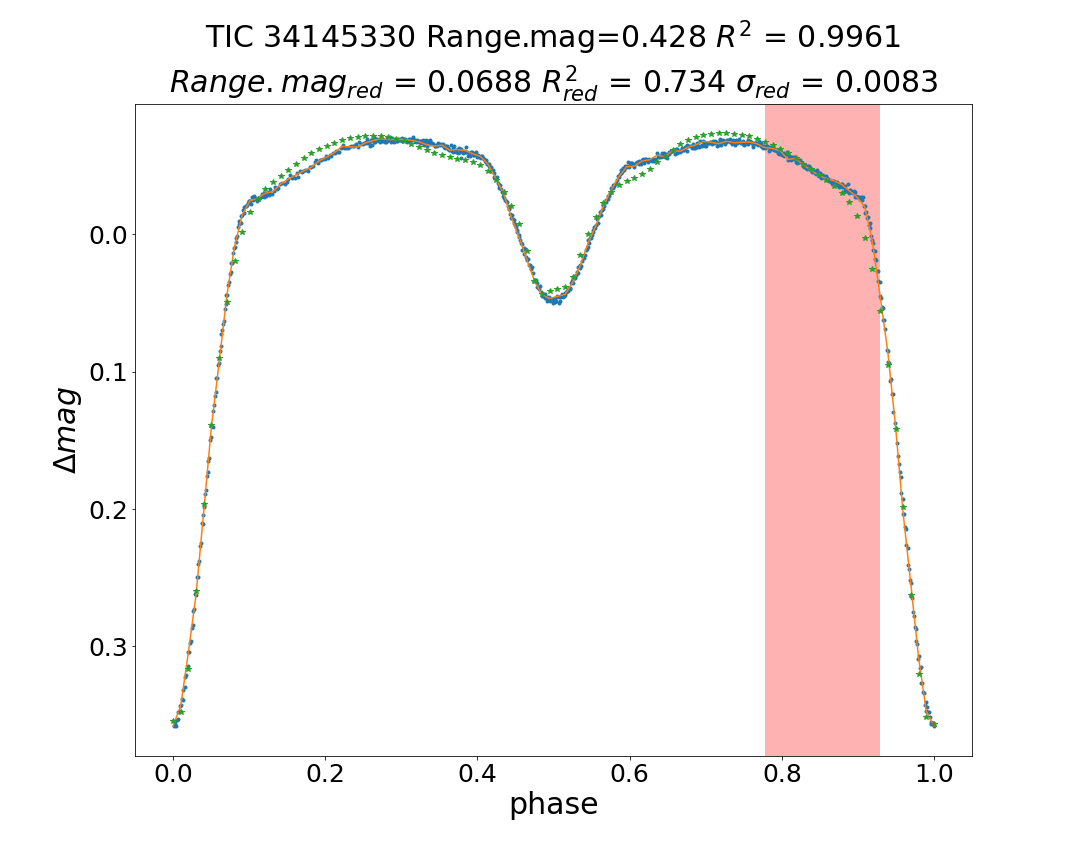}
\end{minipage}
\begin{minipage}{7cm}
	\includegraphics[width=7cm]{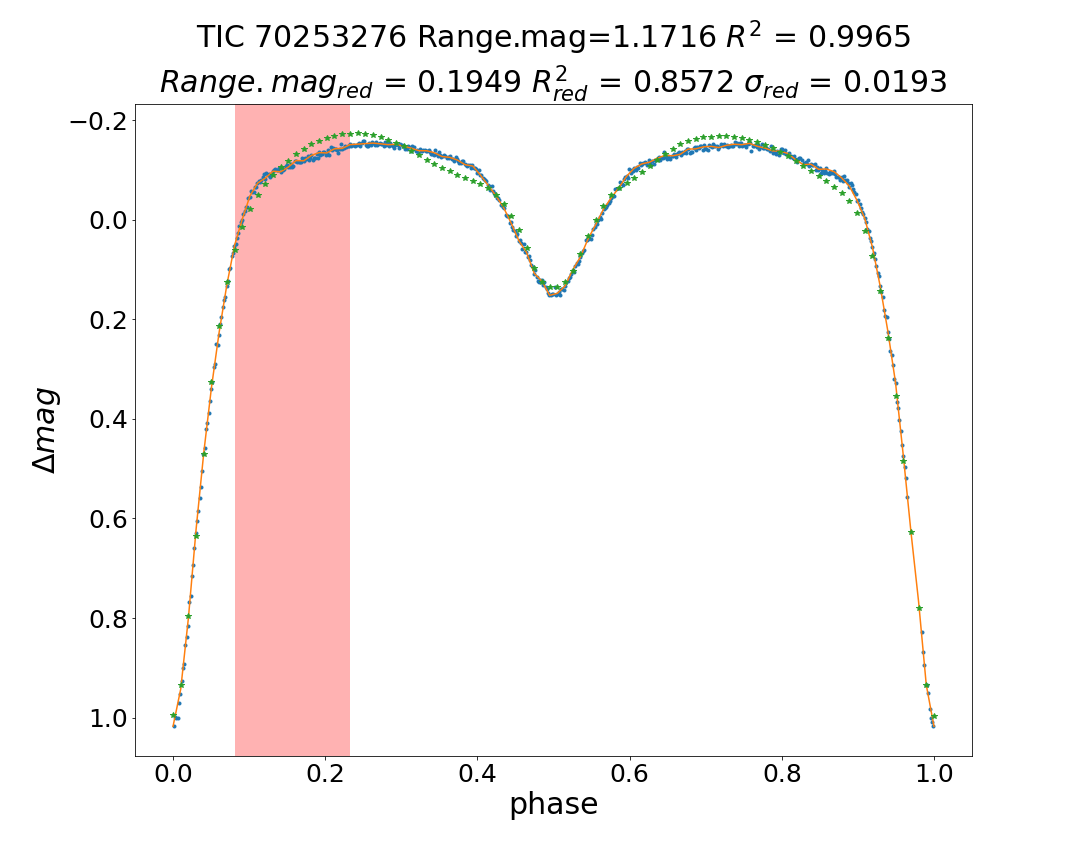}
\end{minipage}
\caption{Similar to Figure 9. The local conditions for sliding window are represented by $Range.mag_{red}$, $R^2_{red}$ and $\sigma_{red}$. $\sigma_{red}$ represents the standard deviation of local residuals. The red bar graph highlights the areas with the poorest local goodness of fit ($R^2$).}\label{Fig99}
\end{center}
\end{figure}

\clearpage
\begin{figure}[!ht]
\begin{center}
\begin{minipage}{6.5cm}
	\includegraphics[width=6.5cm]{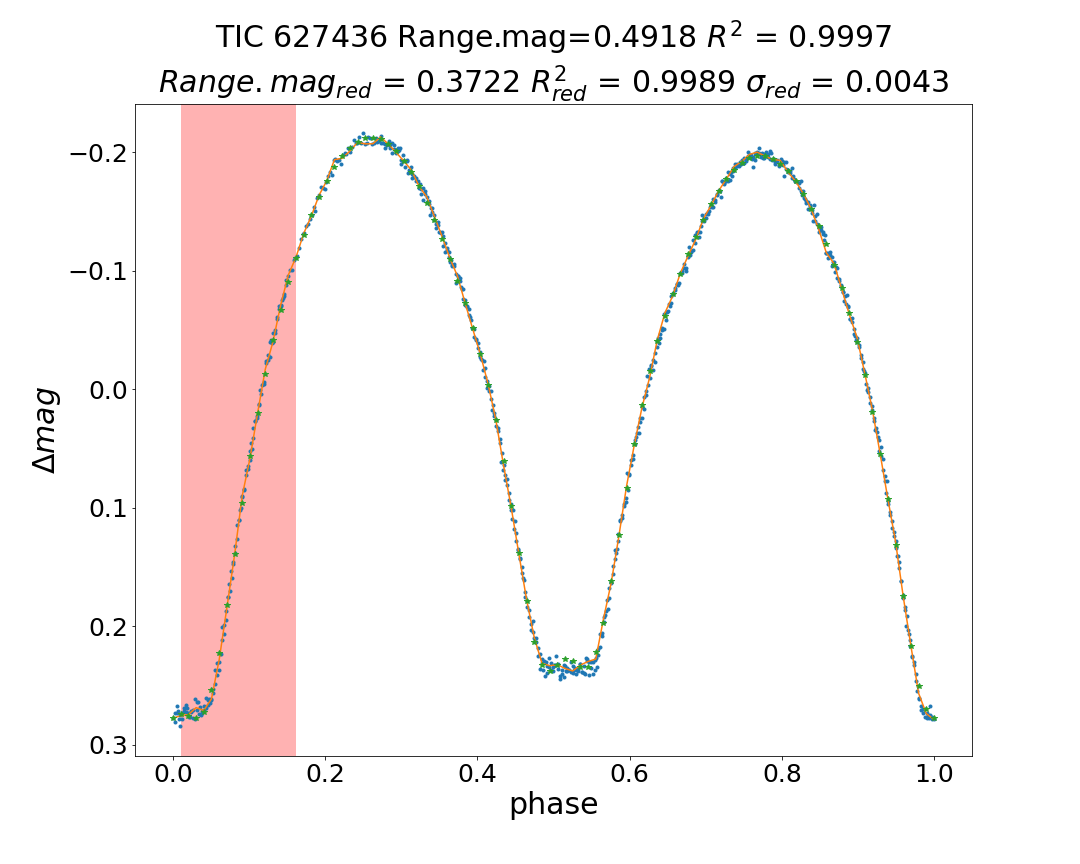}
\end{minipage}
\begin{minipage}{6.5cm}
	\includegraphics[width=6.5cm]{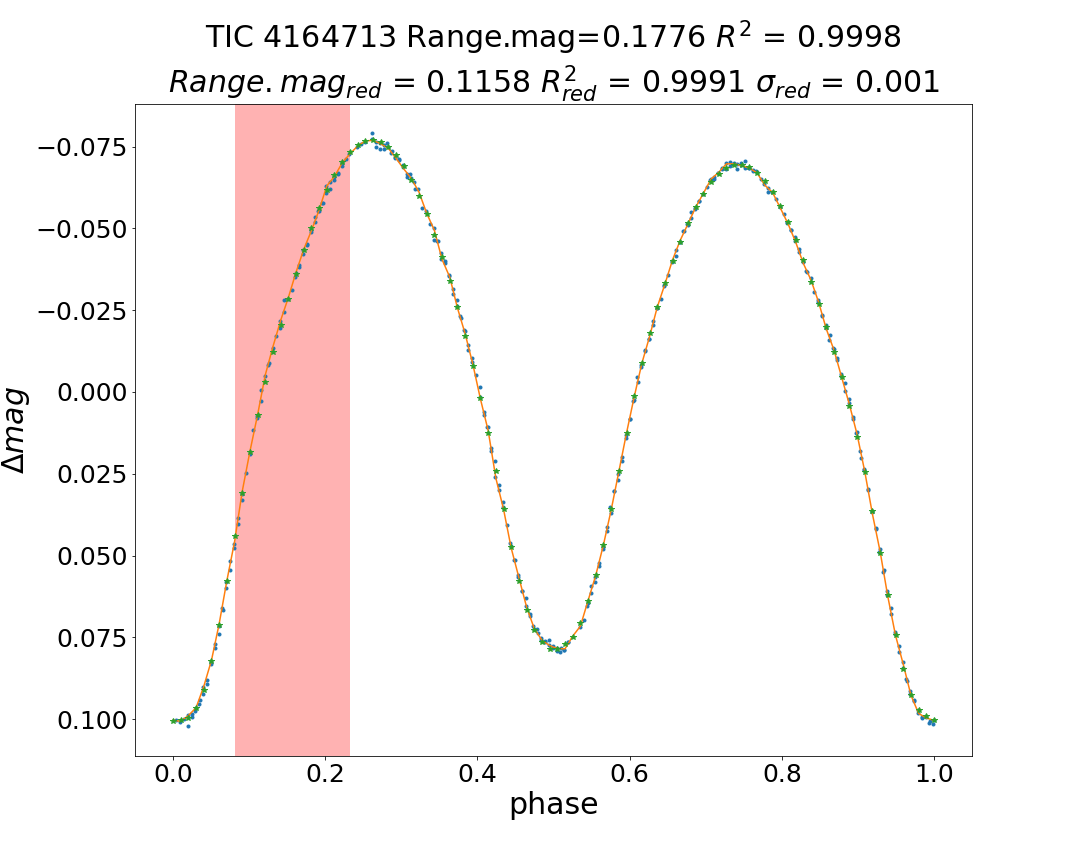}
\end{minipage}
\begin{minipage}{6.5cm}
	\includegraphics[width=6.5cm]{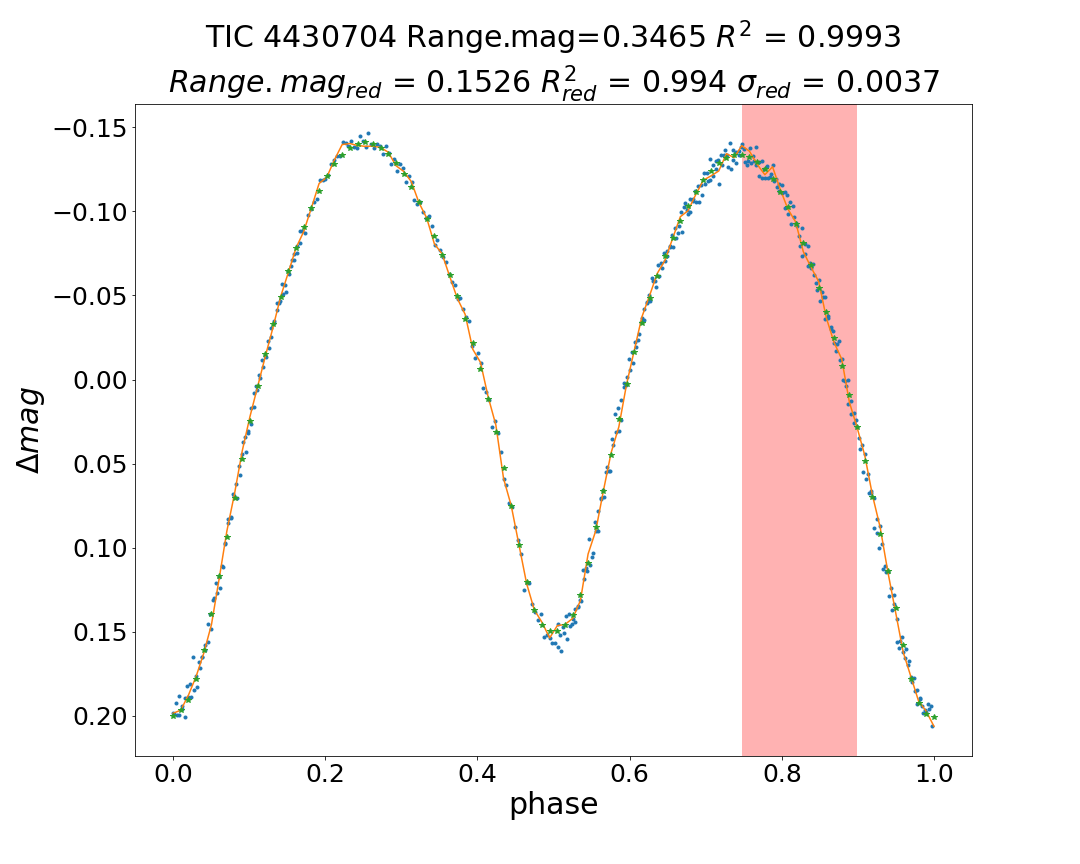}
\end{minipage}
\begin{minipage}{6.5cm}
	\includegraphics[width=6.5cm]{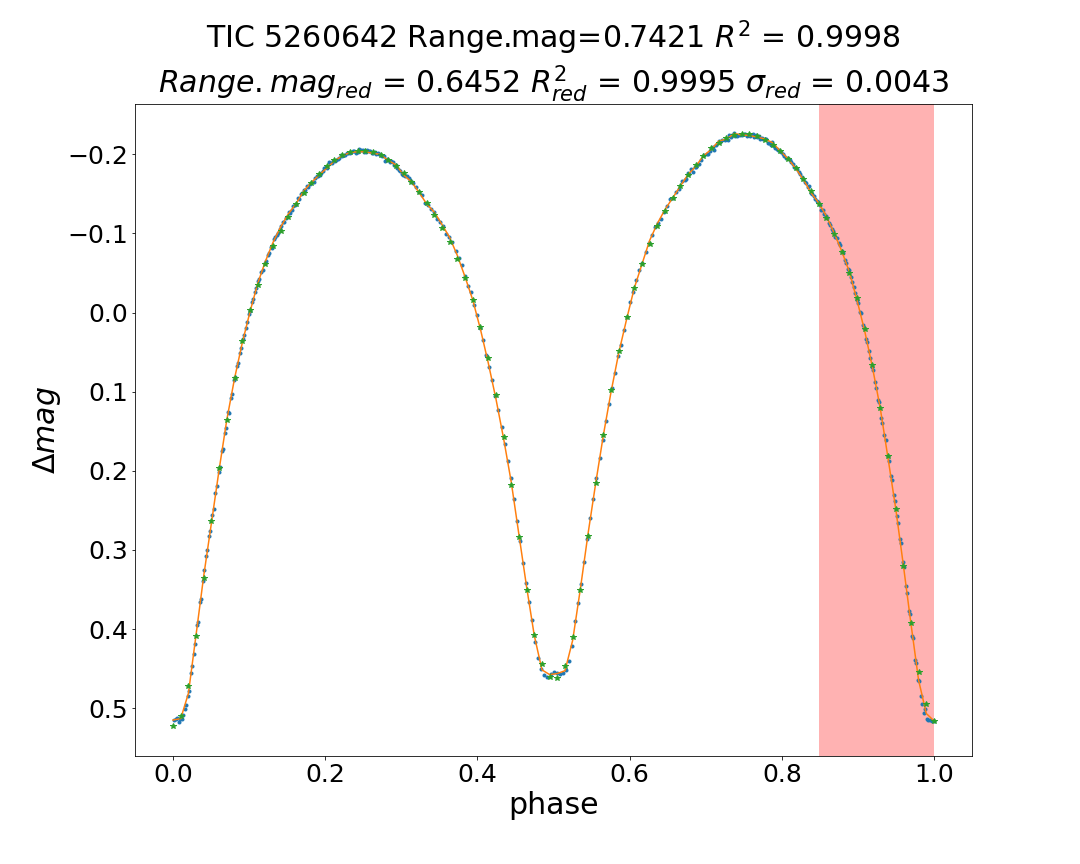}
\end{minipage}
\begin{minipage}{6.5cm}
	\includegraphics[width=6.5cm]{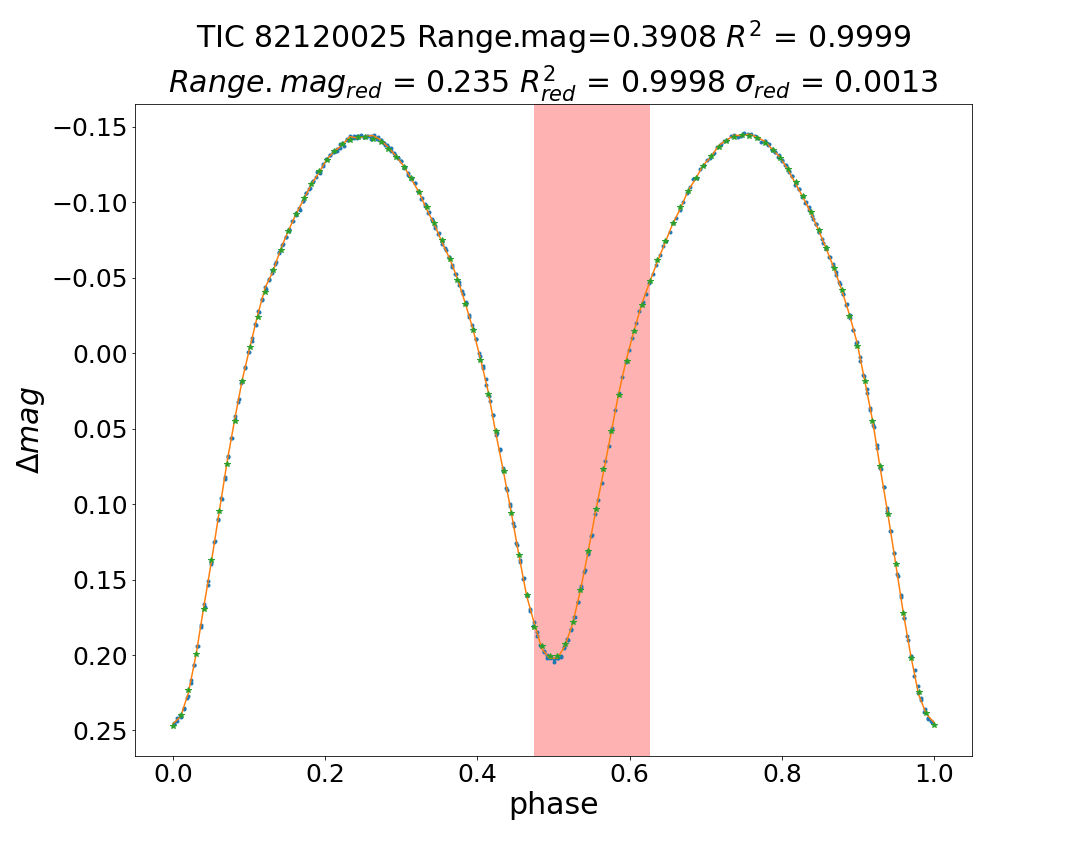}
\end{minipage}
\begin{minipage}{6.5cm}
	\includegraphics[width=6.5cm]{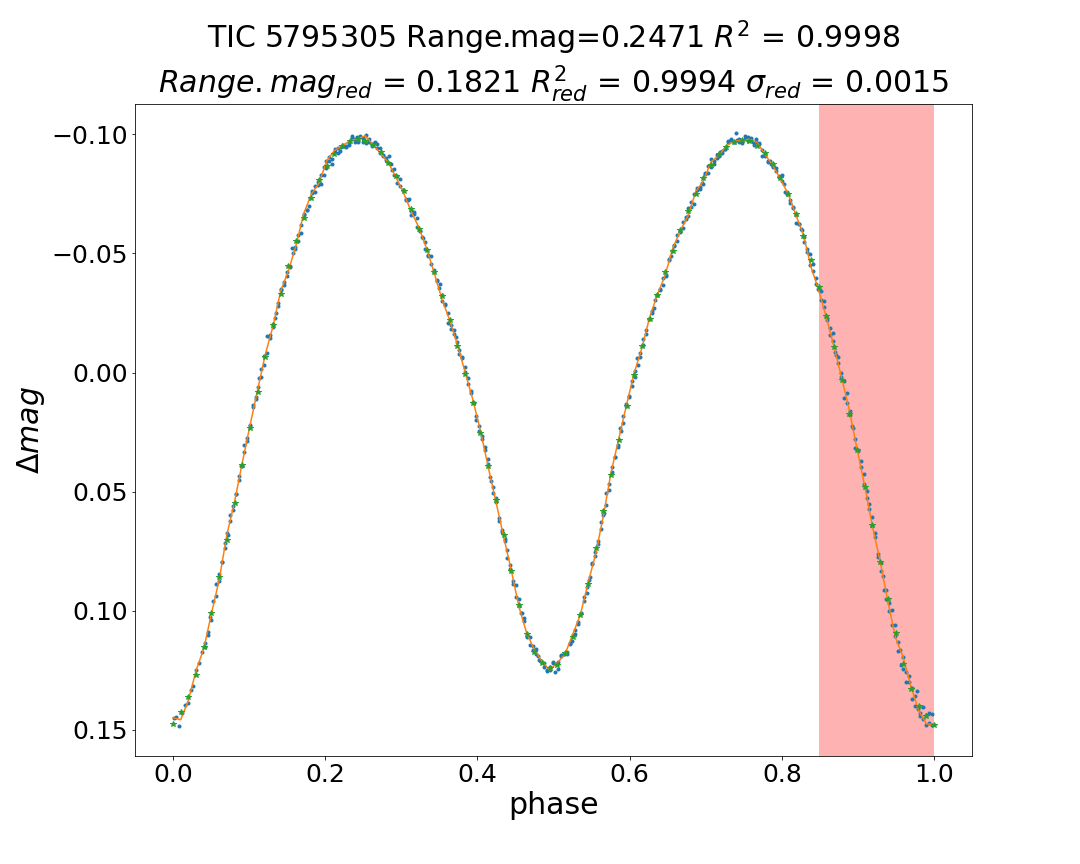}
\end{minipage}
\begin{minipage}{6.5cm}
	\includegraphics[width=6.5cm]{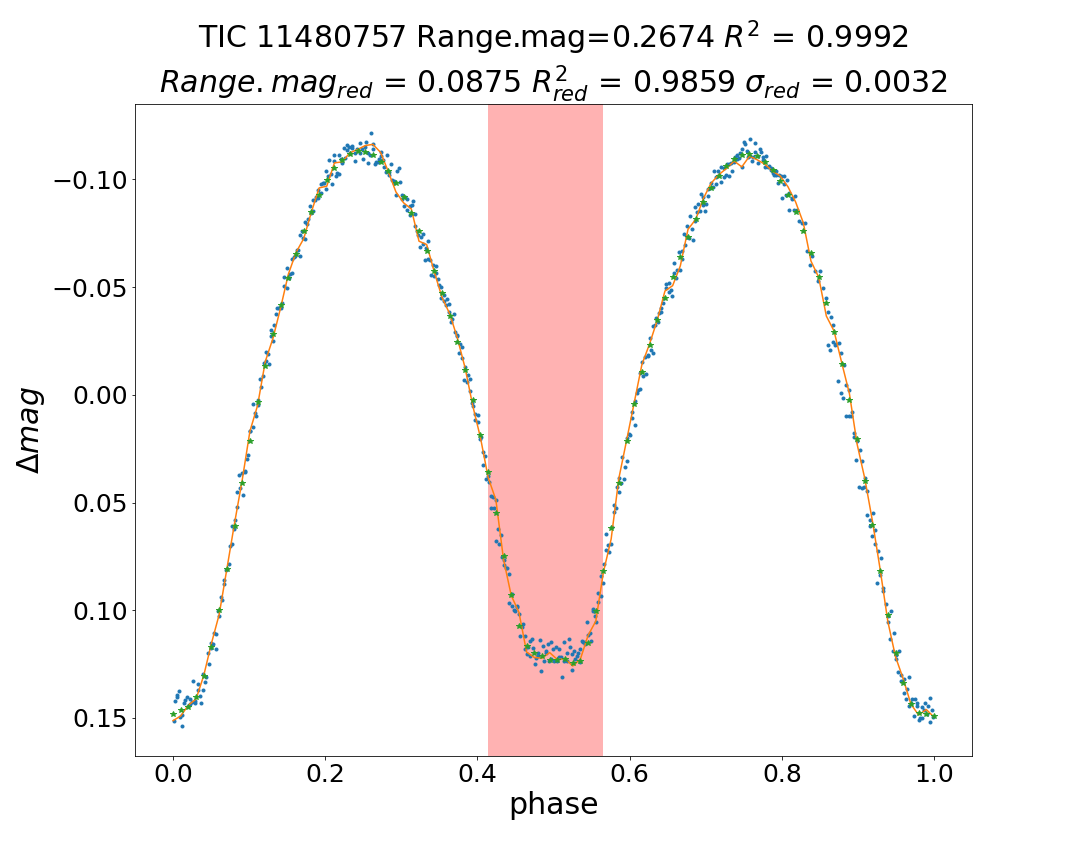}
\end{minipage}
\begin{minipage}{6.5cm}
	\includegraphics[width=6.5cm]{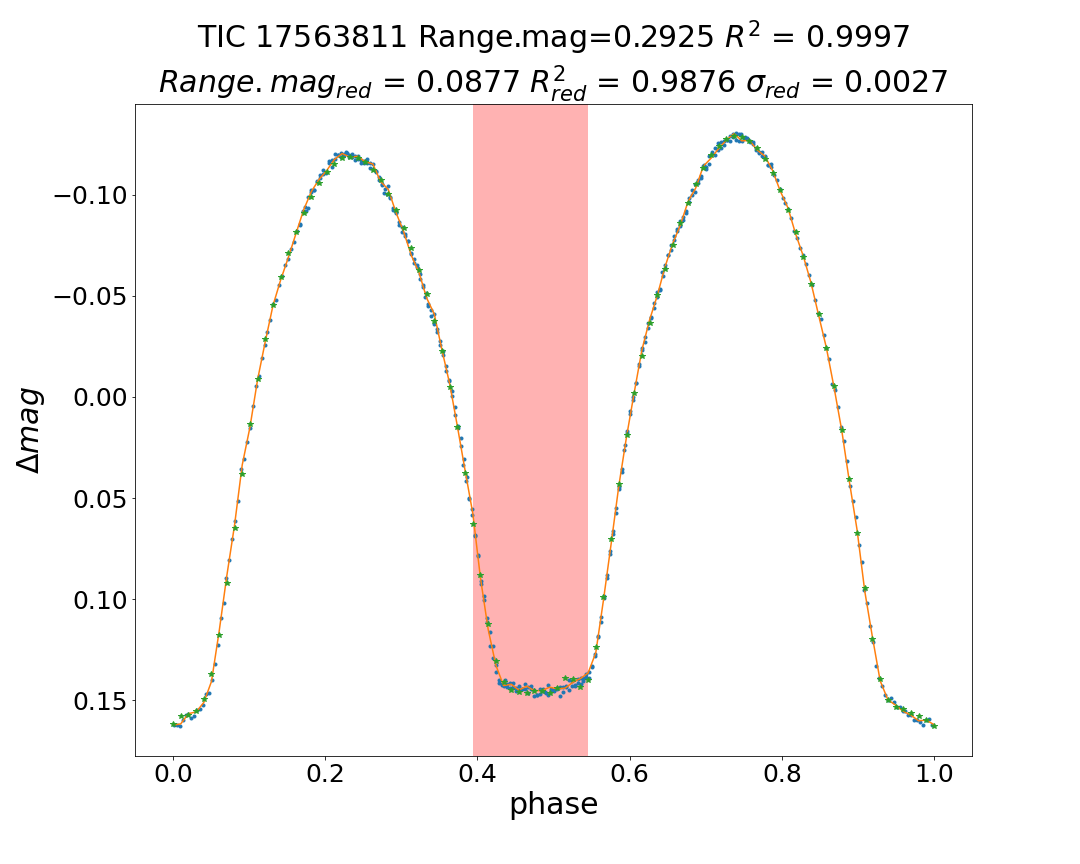}
\end{minipage}

\caption{Similar to Figure 10. The light curves of contact binary candidates can be reconstructed well.}\label{Fig9}
\end{center}
\end{figure}

\begin{figure}[!ht]
\begin{center}
\begin{minipage}{5.5cm}
	\includegraphics[width=5.5cm]{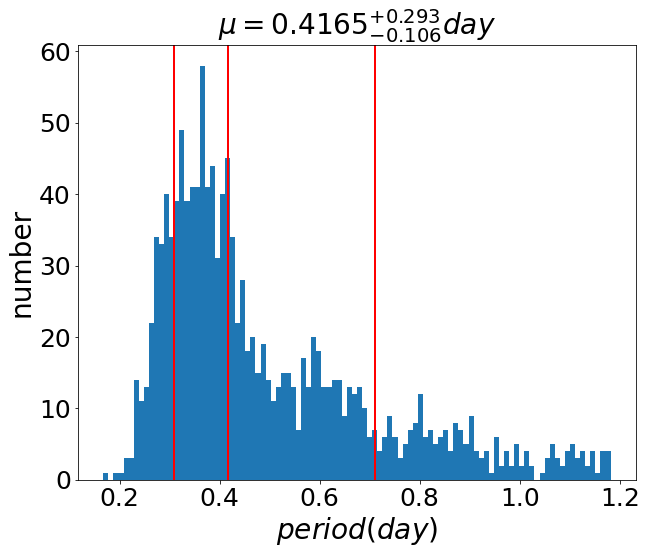}
\end{minipage}
\begin{minipage}{5.5cm}
	\includegraphics[width=5.5cm]{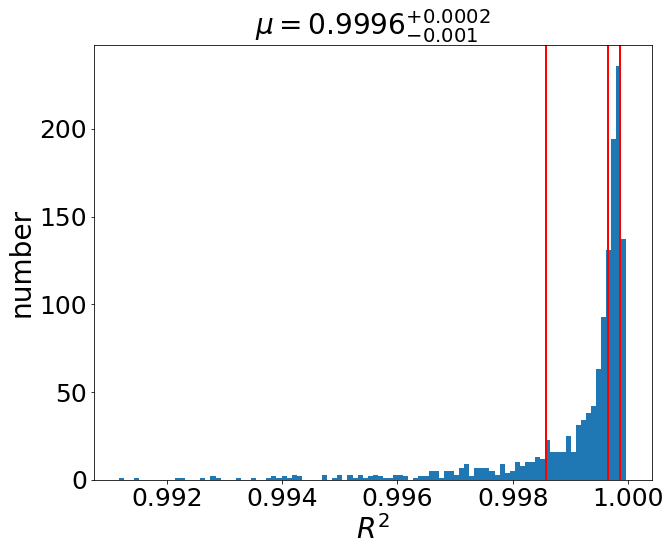}
\end{minipage}
\begin{minipage}{5.5cm}
	\includegraphics[width=5.5cm]{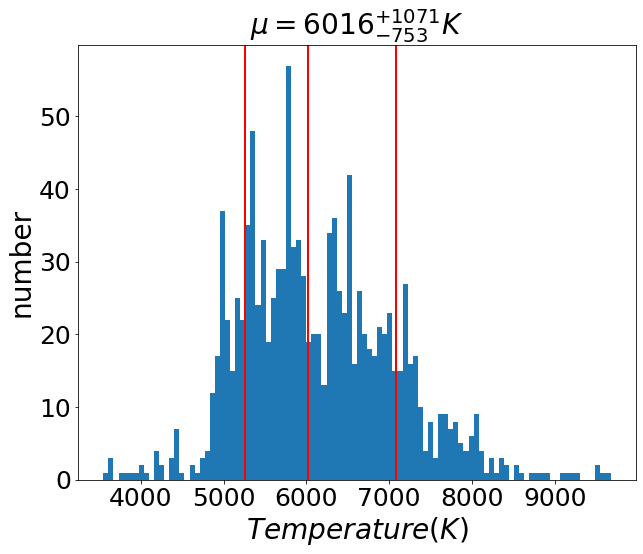}
\end{minipage}
\caption{The period, goodness of fit($R^2$), and temperature distributions are shown.}\label{Fig92}
\end{center}
\end{figure}

\begin{figure}[!ht]
\begin{center}
\begin{minipage}{10cm}
	\includegraphics[width=10cm]{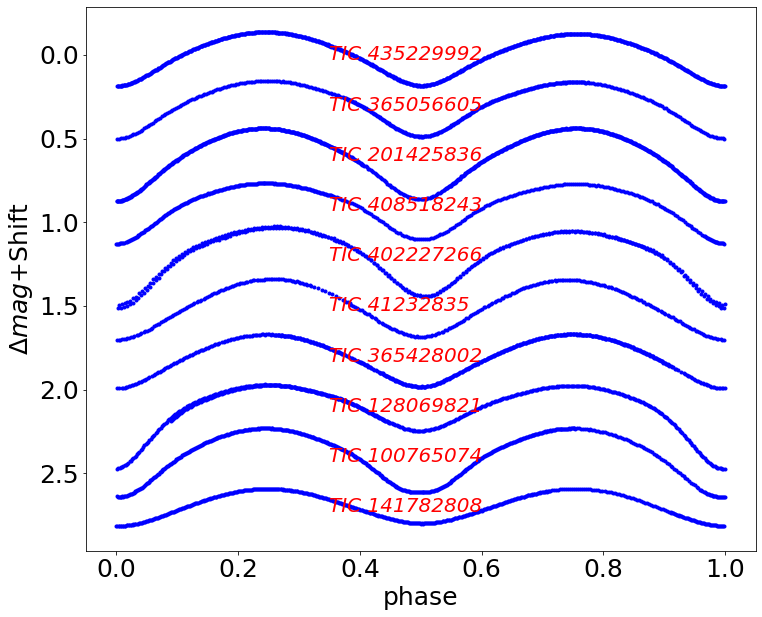}
\end{minipage}
\caption{The examples of 10 candidate objects of contact binary stars that are not present in catalog obtained by \citet{prsa+et+al+2022}.}\label{Fig912}
\end{center}
\end{figure}

\clearpage

\subsection{Two new candidates with ultrashort period}

There are two new candidates, TIC 149917543 and TIC 157365951, with a short period of less than 0.2 days. The light curves of these targets are shown in Figure 14. The fundamental information is shown in Table 4. The temperature of the system was obtained by matching it with Gaia DR2 \citep{Gaia+et+al+2018}. TIC 149917543 and TIC 157365951 were first discovered in the ZTF sky survey \citep{Chen+et+al+2020}. \citet{Chen+et+al+2020} classified TIC 149917543 as a variable star. The target type could not be identified due to the low magnitude variation of this target and the low photometric precision observed by the ZTF telescope.  \citet{Chen+et+al+2020} classified TIC 157365951 as a $\delta$ Scuti type star. 

 These target showed a clear eclipse signal. The depths of the primary and secondary minima of these targets are almost equal or differ insignificantly. 
The light curves of these two targets have ellipsoidal variations and were not fully confirmed as contact binaries. We categorize them as contact binary star candidates in this work. These two targets require further spectroscopic observations to determine the radial velocity variation, and then the Phoebe combined MCMC algorithm \citep{Foreman+et+al+2019} can obtained the fundamental parameters. Constraints in spectral observations and physical models can further determined they are whether or not the contact binaries.

{
\tiny
\begin{center}
\begin{longtable}{ccccccc}
\caption{The fundamental information of two new candidates}\label{Table 3}\\
\hline\hline                          
Parameter &TIC 149917543     &TIC 157365951       \\
              
\hline
\endhead
\hline
RA           &160.64492   &  238.270167                 \\
DEC          &64.701506   &  44.962379                  \\
Period(day)  &0.197640    &  0.167565                    \\
$T_{eff}$(K)   &3666.70  &  4065.55                     \\   
\hline
\end{longtable}
\end{center}
}

\begin{figure}[!ht]
\begin{center}
\begin{minipage}{9cm}
	\includegraphics[width=9cm]{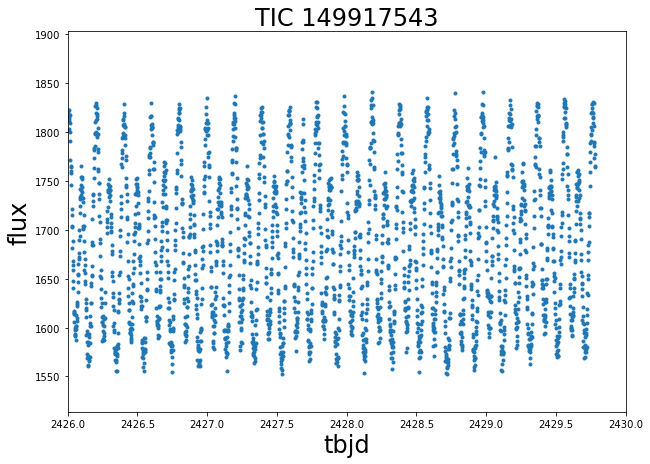}
\end{minipage}
\begin{minipage}{7cm}
	\includegraphics[width=7cm]{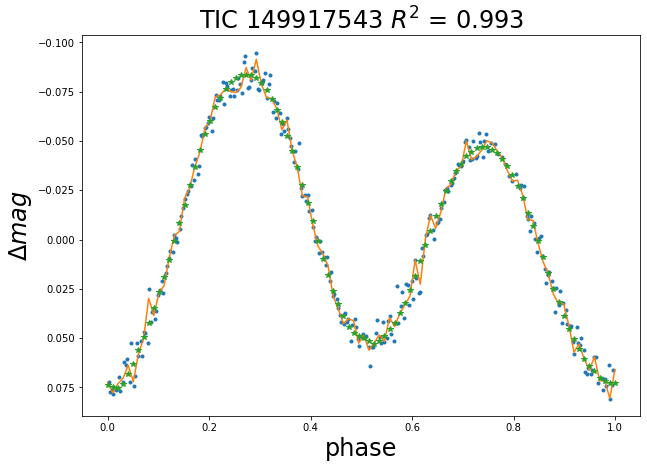}
\end{minipage}
\begin{minipage}{9cm}
	\includegraphics[width=9cm]{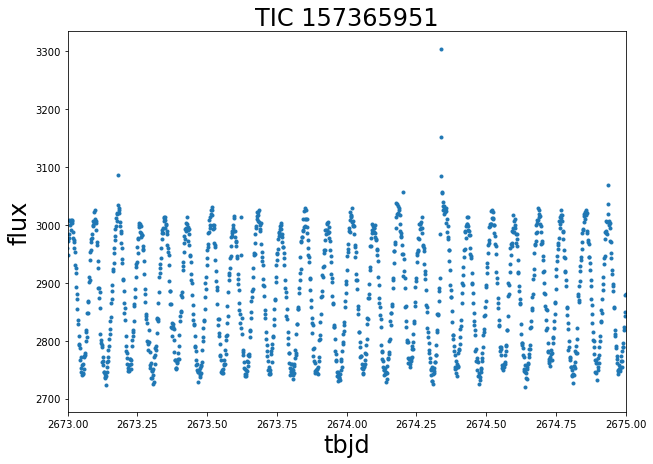}
\end{minipage}
\begin{minipage}{7cm}
	\includegraphics[width=7cm]{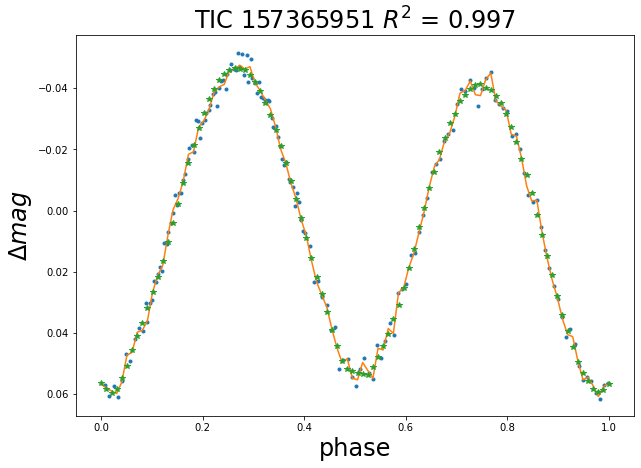}
\end{minipage}
\caption{Left: The blue dots are the light curve with time-flux observed by TESS. Right: Similar to Figure 8 and Figure 9, the blue dots indicate the original light curve, the orange lines indicate the interpolated curves, and the green dots indicate the reconstructed light curves.}\label{Fig14}
\end{center}
\end{figure}

\clearpage{}
\section{Discussion and Conclusions}

With the release of a massive amount of light curves, using supervised machine learning methods \citep{Kang+et+al+2023} for light curve classification requires manual labeling of samples, which incurs significant manual costs. In this work, we use an unsupervised autoencoder neural network to detect contact binary candidates. By utilizing Phoebe, we generate light curve samples of contact binary stars without the need for manual labeling. We take into account an adequate number of parameters to generate the light curves of TESS.
Using just one spot may not fulfill all conditions, but it can satisfy the majority of cases. According to the catalog referenced in \cite{Latkovi+et+al+2021}, as there were fewer literature available in 2021, there were a total of 30 published targets recorded for the year 2020 in Table 5. Among these targets, only the goal named BB Peg stands out as it added two additional spots. Temperature range (M-A spectral type) may not be sufficiently comprehensive, lacking coverage of more early-type spectral contact binaries, such as O and B spectral types. The temperature covers most of the targets for the number of contact binary systems of O and B type is also small.

\clearpage{}
{
\tiny
\begin{center}
\begin{longtable}{ccccccc}
\caption{In the catalog obtained from \cite{Latkovi+et+al+2021}, there are 30 targets published for the year 2020. Only one target, BB Peg , added 2 spots.}\label{Table 23}\\
\hline\hline                          
Name &Bibcode     &Spots Number       \\
              
\hline
\endhead
\hline
1SWASP J003033.05+574347.6           &2020AJ....159..189L   &  1                 \\
1SWASP J161858.05+261303.5           &2020AJ....159..189L   &  1                  \\
ASAS J174406+2446.8                  &2020RAA....20...96S   &  1                  \\
ASAS J212236+0657.3                  &2020BlgAJ..32...71K   &  1                  \\   
ASASSN-V J023009.35+563849.7         &2020MNRAS.497.3381G   &  1                  \\
BB Peg                               &2020NewA...7801354K   &  2                  \\
BF Pav                               &2020arXiv200805068P   &  1                  \\
BQ Ari	                             &2020arXiv200600528P	&  1                  \\
CRTS J130945.0+371627	             &2020AJ....159..189L	&  1                  \\
CRTS J224015.4+184738	             &2020AJ....159..189L	&  1                  \\
GW Leo	                             &2020arXiv201102903P	&  1                  \\
HN Psc	                             &2020NewA...8101439D	&  1                  \\
Mis V1395	                         &2020MNRAS.497.3381G	&  1                  \\
NEV 64	                             &2020MNRAS.497.3381G	&  1                  \\
NSVS 1917038	                     &2020MNRAS.497.3381G	&  1                  \\
NSVS 1926064	                     &2020NewA...7701352K	&  1                  \\
NSVS 4316778	                     &2020NewA...7701352K	&  1                   \\
NSVS 5810460	                     &2020BlgAJ..32...71K	&  1                  \\
NSVS 667994	                         &2020NewA...7701352K	&  1                  \\
PP Lac	                             &2020NewA...7701352K	&  1                   \\
V1071 Per	                         &2020MNRAS.497.3381G	&  1                  \\
V1197 Her	                         &2020RAA....20...10Z	&  1                  \\
V2240 Cyg	                         &2020NewA...7901391O	&  1                  \\
V384 Ser	                         &2020MNRAS.491.6065Z	&  1                  \\
V454 Dra	                         &2020arXiv200209149K	&  1                  \\
V455 Dra	                         &2020arXiv200209149K	&  1                  \\
V599 Aur	                         &2020AJ....160...62H	&  1                   \\
V860 Aur	                         &2020MNRAS.491.6065Z	&  1                   \\
WISE J004327.7+722407	             &2020SerAJ.200...19K	&  1                   \\
WISE J234557.8+510456	             &2020SerAJ.200...19K	&  1                   \\
\hline
\end{longtable}
\end{center}
}

 The autoencoder model has a good reconstruction of the light curves of contact binaries. The distribution of the goodness of fit ($R^2$) between the output light curve from the trained model and the corresponding input light curve was $\sim0.999$. We set the threshold for global goodness of fit ($R^2$) at 0.99, \edit1{the threshold for range magnitude at 0.1}, and the threshold for period at 1.2 days for filtering, which may result in missing some candidate contact binary systems at the critical value range. We further exclude targets with poor local goodness of fit ($R^2$) from the remaining dataset, resulting in a total of 1322 targets obtained. For the phenomenon of short-period cutoff in contact binaries, we have found two new candidates with periods less than 0.2 days, but these candidates need further spectroscopic observations to determine whether they are contact binaries. For contact binaries with extreme mass ratios, a merging event may occur, but the accurate mass ratio determined by Phoebe and Markov chain Monte Carlo (MCMC) is extremely time-consuming. \citet{Ding+et+al+2022} utilized a combination of Markov chain Monte Carlo (MCMC) and neural networks (NN) to rapidly obtain parameters for contact binaries about $\sim30$ seconds on regular computer. In the future, we will employ this method to search for contact binaries with extreme mass ratios using the light curves of candidates obtained from this work.

In order to help other researchers use our code, we will share the code on Github which is consistent with the website address provided in Table 3.

\begin{acknowledgments}
We are very grateful for the data released by the TESS survey (https://archive.stsci.edu/missions-and-data/tess). This work was partly supported by the Chinese Natural Science Foundation (Nos.12103088), Yunnan Provincial Foundation (Nos. 202101AT070020), National Key R\&D Program of China (No.2022YFF0711500), Yunnan Provincial Key Laboratory of Forensic Science  (No. YJXK005) and Yunnan Basic Research Program (No. 202201AU070116). We acknowledge the science research grant from the China Manned Space Project with No.CMS-CSST-2021-A10, No.CMS-CSST-2021-A12 and No.CMS-CSST-2021-B10.

\end{acknowledgments}




\end{document}